\pdfoutput=1

\documentclass{aa}  
\usepackage{graphicx}

\usepackage{subcaption}
\usepackage{natbib}
\usepackage[LGRgreek]{mathastext}
\usepackage{txfonts}
\usepackage{hyperref}

\begin{document}

   \title{The MAGPI survey: The interdependence of the mass, star formation rate, and metallicity in galaxies at $z\sim 0.3$}

   \author{M. Koller\inst{1} \and B. Ziegler\inst{1} \and B. I. Ciocan\inst{1,2} \and S. Thater\inst{1} \and J. T. Mendel\inst{3,4} \and E. Wisnioski\inst{3,4} \and A. J. Battisti\inst{3,4} \and \\K. E. Harborne\inst{4,5} \and C. Foster\inst{4,7} \and C. Lagos\inst{4,5} \and S. M. Croom\inst{4,6} \and  K. Grasha\inst{3,4} \and P. Papaderos\inst{15} \and R. S. Remus\inst{14} \and \\G. Sharma\inst{10,11,12,13} \and S. M. Sweet\inst{4,9} \and L. M. Valenzuela\inst{14} \and G. van de Ven\inst{1} \and T. Zafar\inst{8}}

   \authorrunning{M. Koller et al.}

   \institute{University of Vienna, Department of Astrophysics, 
              T\"urkenschanzstrasse 17, 1180 Vienna, Austria\\
              \email{maria.koller@univie.ac.at}
         \and  Univ Lyon1, CNRS, Centre de Recherche Astrophysique de Lyon, F-69230, Saint-Genis-Laval, France
         \and Research School of Astronomy and Astrophysics, Australian National University, Cotter Road, Weston Creek, ACT 2611, Australia
         \and ARC Centre of Excellence for All Sky Astrophysics in 3 Dimensions (ASTRO 3D)
        \and International Centre for Radio Astronomy (ICRAR), M468, The University of Western Australia, 35 Stirling Highway, Crawley, WA 6009, Australia
        \and Sydney Institute for Astronomy, School of Physics, University of Sydney, NSW 2006, Australia
        \and School of Physics, University of New South Wales, Sydney, NSW 2052, Australia
        \and School of Mathematical and Physical Sciences, Macquarie University, NSW 2109, Australia
        \and School of Mathematics and Physics, University of Queensland, Brisbane, QLD 4072, Australia
         \and University of Strasbourg, CNRS UMR 7550, Observatoire astronomique de Strasbourg, F-67000 Strasbourg, France
         \and University of Strasbourg Institute for Advanced Study, 5 allée du Général Rouvillois, F-67083 Strasbourg, Franc
         \and SISSA International School for Advanced Studies, Via Bonomea 265, I-34136 Trieste, Italy
         \and Department of Physics and Astronomy, University of the Western Cape, Cape Town 7535, South Africa
         \and Universitäts-Sternwarte, Fakultät für Physik, Ludwig-Maximilians-Universität München, Scheinerstr. 1, 81679 München, Germany
         \and Instituto de Astrofísica e Ciências do Espaço - Centro de Astrofísica da Universidade do Porto, Rua das Estrelas, 4150-762 Porto, Portugal
            }

   \date{Received XXX /
    Accepted XXX }

% \abstract{}{}{}{}{} 
% 5 {} token are mandatory
 
  \abstract
  % context heading (optional)
  % {} leave it empty if necessary  
   {}
  % aims heading (mandatory)
   {Star formation rates (SFRs), gas-phase metallicities, and stellar masses are crucial for studying galaxy evolution. The different relations resulting from these properties give insights into the complex interplay of gas inside galaxies and their evolutionary trajectory and current characteristics. We aim to characterize these relations at $z\sim 0.3$, corresponding to a 3-4 Gyr lookback time, to gather insight into the galaxies' redshift evolution.}
  % methods heading (mandatory)
   {We utilized optical integral field spectroscopy data from 65 emission-line galaxies from the MUSE large program MAGPI at a redshift of $0.28<z<0.35$ (average redshift of $z\sim 0.3$) and spanning a total stellar mass range of $8.2<\log(M_{*}/M_{\odot}) < 11.4$. We measured emission line fluxes and stellar masses, allowing us to determine spatially resolved SFRs, gas-phase metallicities, and stellar mass surface densities. We derived the resolved star formation main sequence (rSFMS), resolved mass metallicity relation (rMZR), and resolved fundamental metallicity relation (rFMR) at $z\sim 0.3$, and compared them to results for the local Universe.}
  % results heading (mandatory)
   {We find a relatively shallow rSFMS slope of $\sim 0.425 \pm 0.014$ compared to the expected slope at this redshift for an ordinary least square (OLS) fitting routine. For an orthogonal distance regression (ODR) routine, a much steeper slope of $\sim 1.162 \pm 0.022$ is measured. We confirm the existence of an rMZR at $z\sim 0.3$ with an average metallicity located $\sim 0.03$ dex above the local Universe's metallicity. Via partial correlation coefficients, evidence is found that the local metallicity is predominantly determined by the stellar mass surface density and has a weak secondary (inverse) dependence on the SFR surface density $\Sigma_{SFR}$. Additionally, a significant dependence of the local metallicity on the total stellar mass $M_{*}$ is found. Furthermore, we find that the stellar mass surface density $\Sigma_{*}$ and $M_{*}$ have a significant influence in determining the strength with which $\Sigma_{SFR}$ correlates with the local metallicity. We observe that at lower stellar masses, there is a tighter correlation between $\Sigma_{SFR}$ and the gas-phase metallicity, resulting in a more pronounced rFMR.}
  % conclusions heading (optional), leave it empty if necessary 
   {}


   \maketitle
%
%-------------------------------------------------------------------

\section{Introduction}\label{sec:intro}

The processes governing galaxy evolution set the local and global gas-phase chemical abundances and stellar masses of galaxies \citep{Madau_2014, Lilly_2013}. The rate of inflowing and outflowing gas, feedback processes such as stellar winds and supernovae (SNe), and the rate at which gas is recycled within galaxies determine the abundance of metals, measured as the gas-phase metallicity, and star formation activity \citep{Peroux_2020}. Gas-phase metallicity is traced via strong ionized gas emission lines \citep{Kewley_2019}, which are correlated with the amount of metals produced in past stellar generations \citep{Maiolino_2019_deremet}. Inflowing pristine (metal-poor) gas from the intergalactic medium (IGM) both dilutes the overall gas content and increases the star formation rate (SFR) \citep{Tumlinson_2017}. Stellar evolution plays a role in this gas cycle by providing chemically enriched gas to the interstellar medium (ISM) through SN explosions and stellar winds. High metallicity gas can leave the galaxy through galactic winds to enrich the circumgalactic medium (CGM). The gas inside galaxies is recycled multiple times and undergoes numerous stellar generations. Dust is created and destroyed, which captures and releases metals. The presence of a central active galactic nucleus (AGN) impacts the properties of the ISM and the CGM, either by heating the gas or by removing it through AGN-driven winds, and both of these effects can impede star formation \citep{Lilly_2013, Peng_Maiolino_2014}.

This interdependence between ionized gas properties has been studied for decades. It is essential to investigate to better understand the cosmic baryon cycle, which defines the star formation activity and metal content present in the gas of galaxies \citep{Peroux_2020}. One of these relations is the correlation between gas-phase metallicity and stellar mass, the so-called mass-metallicity relation (MZR, see e.g., \citealt{Tremonti_2004, Kewley_Ellison_2008_MZR}; for a review \citealt{Maiolino_2019_deremet}). This correlation was first defined as a metallicity versus luminosity relation by \cite{Lequeux_1979}. An increase in stellar mass correlates with an increase in metallicity until the relation flattens out at high stellar masses. \cite{Tremonti_2004} studied this relation using 53,000 star-forming galaxies at $z=0.1$ from the Sloan Digital Sky Survey (SDSS; \citealt{Kollmeier_2019}) and found that the MZR has a $\sim 0.1$ dex scatter spanning three orders of magnitude in stellar mass and a factor of ten in metallicity. This relation has been studied for the local Universe and redshifts up to $z \sim 10$ \citep{Maiolino_2008, Lamareille_2009_MZR, Mannucci_2010, Maier_2014_MZR, Cresci_2019_FMR, Curti_2019_MZR, Curti_2022, Nakajima_2023_FMR}. An evolution with redshift has been observed: high redshift galaxies are more metal-poor than local galaxies \citep{Tremonti_2004, Savaglio_2005_MZR, Maiolino_2008, Gao_2018_FMR_highz, Sanders_2021_MZR}. The MZR's redshift evolution is commonly attributed to galaxies in the local Universe having had a longer time to produce and assemble metals than galaxies at high redshifts. 

Another correlation is established between the SFR and the stellar mass of a galaxy: the star formation main sequence (SFMS, see e.g., \citealt{Brinchmann_2004_SFMS, Noeske_2007_SFMS, Whitaker_2012, Speagle_2014_SFMS, Renzini_2015_SFMS}). Star-forming galaxies inhabit a near-linear relation in the SFR versus stellar mass space. Galaxies that experience star formation enhancements or suppression are situated above or below this line, respectively \citep{Noeske_2007_SFMS}. The SFMS is also observed to evolve with redshift: galaxies at a given $M_{*}$ have higher SFRs at higher redshifts \citep{Speagle_2014_SFMS, Pearson_2018, Leslie_2020}. 

Connecting all three properties (stellar mass, SFR, and gas-phase metallicity) establishes a three-dimensional relation: the fundamental metallicity relation (FMR, see e.g., \citealt{Mannucci_2010, Lara-Lopez_2010_FMR,Cresci_2019_FMR, Curti_2019_MZR, Pistis_2023,Li_2024, Curti_2024_FMR}), which is typically not observed to evolve with redshift. For a given stellar mass, galaxies with higher SFRs exhibit a lower metallicity. \cite{Ellison_2008_FMR}, who first discovered this anti-correlation, used over 40,000 galaxies from the SDSS survey to investigate the correlation between the specific star formation rate (sSFR), which is defined as $sSFR = SFR/M_{*}$, and the gas-phase metallicity. They concluded that galaxies with a higher sSFR have systematically lower metallicities of up to 0.2 dex at a fixed stellar mass. Their study also discussed some possible origins for this relationship between the SFR and metallicity, including environmental effects, star formation efficiencies, infall of metal-poor gas, and galactic winds. This anticorrelation was later confirmed by \cite{Mannucci_2010}, who investigated the relationship between the SFR, gas-phase metallicities, and stellar masses of a large sample of galaxies ranging over redshifts of $z=0-2.5$, and \cite{Lara-Lopez_2010_FMR}, who also conducted a study of these three properties for star-forming galaxies of the SDSS survey over a redshift range of $0.04<z<0.1$. Both studies concluded that the galaxies populate a three-dimensional plane, the FMR. \cite{Mannucci_2010} observations also show that the relation is more robust for low-mass galaxies and that the metallicity ceases to depend on the SFR at high stellar masses. 

In recent years, the FMR has been a great topic of discussion. Publications such as \cite{Sanchez_2013_rMZR} and \cite{Barrera_Ballesteros_2016_rMZR} have questioned the relation's existence. Support has been found in the works of \cite{Lara-Lopez_2010_FMR, Mannucci_2010, Salim_2014_FMR}. Currently, the largest studies of the FMR have been done using local galaxies, but investigations at higher redshifts have been made \citep{Cresci_2012_FMR, Cresci_2019_FMR, Curti_2022, Nakajima_2023_FMR, Curti_2024_FMR}. Studies toward higher redshifts where the S/N and overall spatial resolution are much lower than for local galaxies are still scarce. With the advent of the most recent generation of NIR IFU spectrographs, such as VLT/ERIS and JWST/NIRSpec, and more extensive surveys conducted via optical IFU spectrographs, such as VLT/MUSE, this gap may now be addressed. Some studies in the IR using early-release JWST/NIRSpec data have already been performed. For example, \cite{Nakajima_2023_FMR} investigated the FMR and redshift evolution of the MZR at $z=4-10$ from 135 galaxies using JWST/NIRSpec data. They find that only a small MZR evolution is observed from $z=2-3$ to $z=4-10$ while the FMR shows no significant evolution up to $z\sim 8$.

\cite{Cresci_2012_FMR, Cresci_2019_FMR} also studied the redshift evolution of the FMR and find an absence of changes over cosmic time. On the contrary, \cite{Garcia_2024_FMR} studied the FMR via Illustris \citep{Vogelsberger_2014_Illustris}, Illustris The Next Generation (IllustrisTNG; \citealt{Pillepich_2018_IllustrisTNG}), and Evolution and Assembly of GaLaxies and their Environment (EAGLE; \citealt{Schaye_2015_EAGLE}) simulations. Their results are consistent with a "weak" FMR with a non-negligible redshift evolution.

In recent years, spatially resolved spectroscopy of galaxies from surveys of the local Universe (for a review see \citealt{Sanchez_2020}) such as the Calar Alto Legacy Integral Field Area (CALIFA; \citealt{Sanchez_2012_CALIFA}) survey, Mapping Nearby Galaxies at APO (MaNGA; \citealt{Bundy_2015_MaNGA}) survey, and Sydney-AAO Multiobject Integral-field spectrograph (SAMI; \citealt{Croom_2021_SAMI}) survey have become more accessible for larger samples, making it possible to study these relations on kpc-scales. Resolved relations can be established by substituting the measured parameters for surface densities. The resolved star formation main sequence (rSFMS, see e.g., \citealt{Cano_Diaz_2016_rSFMS, Ellison_2018_rSFMS,Medling_2018_rSFMS, Jafariyazani_2019_rSFMS, Yao_2022, Baker_2022_rSFMS_rSK_rMGMS}) describes the relation between stellar mass surface density, $\Sigma_{*}$, and SFR surface density, $\Sigma_{SFR}$. Similarly, a resolved mass metallicity relation (rMZR) has been observed as well by relating $\Sigma_{*}$ and the local metallicity \citep{Rosales_Ortega_2012, Sanchez_2013_rMZR, Barrera_Ballesteros_2016_rMZR, Yao_2022}. Studies by \cite{Sanchez-Menguiano_2019_rFMR, Baker2022_metallicity}, and \cite{Li_2024} have found evidence for a resolved FMR. Nonetheless, clear evidence for the existence of an rFMR even at $z\sim 0$, where the most extensive IFU studies have been conducted so far, has yet to be found (e.g., \citealt{Barrera_Ballesteros_2016_rMZR}). 

Spatially resolved investigations at intermediate redshifts have been conducted \citep{Jafariyazani_2019_rSFMS, Yao_2022}. For instance, \cite{Yao_2022} analyzed the spatially resolved MUSE data of ten star-forming galaxies at $z\sim 0.26$. They find consistent results for an rMZR and rSFMS, but no clear evidence for an rFMR.

More extensive spatially resolved studies toward higher redshifts are needed to get a clear picture of the rMZR's redshift evolution and to find evidence of whether or not an rFMR exists at intermediate redshifts around $ z \sim 0.3$ (3-4 Gyr lookback time). In this context, we aim to bridge this gap between the local Universe and higher redshifts by investigating galaxies spatially resolved ionized gas relations via the Middle Ages Galaxy Properties with Integral field spectroscopy (MAGPI; \citealt{Foster_2021_MAGPI}) survey\footnote{Based on observations obtained at the Very Large Telescope (VLT) of the European Southern Observatory (ESO), Paranal, Chile (ESO program ID 1104.B-0536)} at a 3-4 Gyr lookback time between cosmic noon and the local Universe. 

This paper is structured as follows. In section~\ref{sec:data}, we discuss the survey and data products used in this work and describe our sample selection criteria. In section~\ref{sec:methods}, we discuss how we measure our physical quantities. Our results are presented in section~\ref{sec:results} and discussed in section~\ref{sec:discussion}. Finally, we summarize and conclude our results in section~\ref{sec:conclusion}. Throughout this paper, we assume a \cite{Chabrier_2003_IMF} initial mass function (IMF) and adopt a flat $\Lambda$CDM cosmology with $H_0 = 70.0 \ km \ s^{-1} \ Mpc^{-1}$, $\Omega_m = 0.3$, and $\Omega_\lambda = 0.7$.

\section{Data}\label{sec:data}
\subsection{The MAGPI survey}

The MAGPI survey \citep{Foster_2021_MAGPI} is a VLT/MUSE large program (Program ID: 1104.B-0536) targeting galaxy environments at redshifts $0.28<z<0.35$, corresponding to 3-4 Gyr lookback times. MAGPI's primary goal is to obtain spatially resolved spectroscopic properties of the galaxies' stars and ionized gas to explore the different mechanisms shaping the morpho-kinematics of today's massive galaxies at a crucial time in their evolution. Representing an extension to already existing low-redshift IFU surveys like SAMI and MANGA, MAGPI provides comparable spatial resolutions but at nearly twice the lookback time, thus filling a gap at this intermediate redshift range. 

The survey targets 60 primary galaxies, which were selected based on a stellar mass of $M_{*}>7 \times 10^{10} M_{\odot}$, and $\sim100$ satellites in a wide range of environments. In total, 56 out of the 60 primary targets were selected from the fields G12, G15, and G23 stemming from the Galaxy and Mass Assembly (GAMA, \citealt{Driver_2011_GAMA}) survey. The other four outstanding primary targets were taken from archival MUSE observations of Abell 370 and Abell 2744 fields. 

The observations are conducted with the MUSE Wide Field Mode within a wavelength range of $4650\AA - 9300\AA$ and a spectral sampling of $1.25 \AA$. Each MAGPI field covers a field-of-view of roughly $1^{\prime}x1^{\prime}$ and is centered on a primary target. A spatial sampling of $0.2^{\prime\prime}$ per pixel and an average image quality of $0.65^{\prime\prime}$ FWHM in the V-band is achieved \citep{Foster_2021_MAGPI}. Each field is observed over six observing blocks with an exposure time of $2x1320$s each, resulting in an integrated time of 4.4h per field. Ground-layer adaptive optics (GLAO) alleviates atmospheric seeing, resulting in a $270\AA$ wide gap within the galaxies observed spectra at $5780-6050 \AA$ caused by the GALACSI sodium laser notch filter. A detailed explanation of the data reduction and various spatially resolved map creations are discussed in Mendel et al. (in prep). Currently, observations for MAGPI are ongoing, with 42 out of 60 fields ($\sim 70\%$ completion rate) having been observed and fully reduced. 

For our galaxy selection, we utilized an emission line catalog of the integrated galaxy spectra of all observed targets. This emission line product is fully described in Battisti et al. (in prep). Emission lines are measured via GIST \citep{Bittner_2019_GIST}, which is a Python wrapper for pPXF \citep{Cappellari_2004_ppxf, Cappellari_2017_ppxf} and GANDALF \citep{Sarzi_2017_GANDALF}. For more details on the MAGPI-derived emission line fluxes, see also \cite{Mun_2024}. Morphological parameters are derived from pseudo-i-band images produced from the MUSE data. Half light radii $R_e$ and semi-major axis containing $90\%$ of the flux $R_{90}$ are measured using the Profound R Package \citep{Profound_Robotham_2018}. The total stellar masses $M_*$ used to introduce our sample in Fig.~\ref{fig:sample} were computed via ProSpect \citep{Robotham_2020_prospec}. We also use MAGPI's spectroscopic redshifts derived via MARZ \citep{Hinton_2016_Marz}.

\subsection{Selection of galaxies and spaxels}

This work focuses on a sample of galaxies selected based on several criteria. First, galaxies must lie within MAGPI's primary redshift range ($0.28<z<0.35$), which results in 393 available galaxies. Secondly, we select galaxies based on the S/N of their integrated spectrum of the following emission lines: $H\beta$, $[OIII]\lambda 5007$, $[NII]\lambda 6583$, and $H\alpha$. Galaxies must have a S/N>5 in all four emission lines. Thirdly, galaxies must have an effective radius $R_e$ in i-band of $R_e>0.7^{\prime\prime}$, making it $0.05^{\prime\prime}$ larger than the spatial resolution of the MUSE observations. Fourthly, the galaxy should not be at the edge of the MUSE field or be cut off. Lastly, we remove highly inclined galaxies with a minor ($b^{\prime}$) to major ($a^{\prime}$) axis ratio of $b^{\prime}/a^{\prime}<0.35$ to avoid strong inclination and extreme extinction corrections. Following all the selection criteria, we are left with 65 galaxies. 

\begin{figure}
    \centering
    \includegraphics[width=0.8\linewidth]{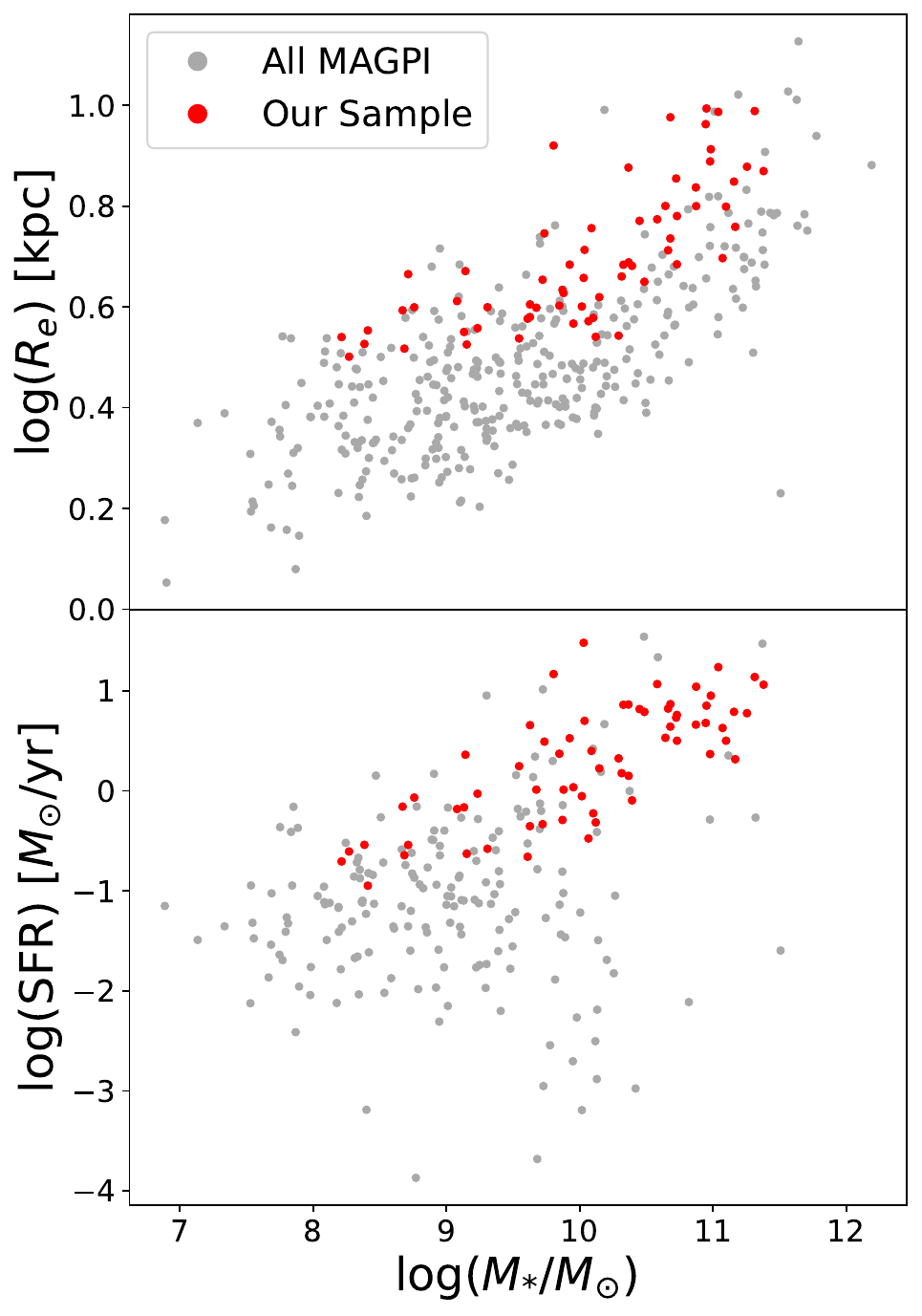}
    \caption{Distribution of our selected sample. \textit{Top:} Mass-size diagram for our selected galaxies (red) vs. the entire MAGPI catalog within the same redshift range of $0.28<z<0.35$ as us (gray). Stellar masses were taken from MAGPI's ProSpect catalog and effective radii from MAGPI's Profound catalog. \textit{Bottom:} Mass-SFR diagram via SFRs derived from integrated $H\alpha$ fluxes taken from MAGPI's GIST measurements.}
    \label{fig:sample}
\end{figure}

In Fig.~\ref{fig:sample}, we show the distribution of our selected sample as mass-size and mass-SFR diagrams and compare them to the entire MAGPI sample within the same redshift range. The star formation rates shown here were computed only for galaxies with $H\alpha$ measurements ($z<0.42$ for the MUSE spectral window) following the equations described in section~\ref{sec:methods_SFR} and corrected for intrinsic extinction as described in section~\ref{sec:fluexes_meth}. The $H\alpha$ flux measurements were taken from MAGPI's integrated emission line catalog using GIST (Battisti et al. in prep). 

We extracted spaxels from the MAGPI cubes via the MUSE Python Data Analysis Framework (MPDAF;  \citealt{Bacon_2016_MPDAF}) and corrected for Galactic foreground extinction using the \texttt{noao onedspec deredden} routine from the Image Reduction and Analysis Facility (IRAF; \citealt{Tody_1986_IRAF}), which is based on the empirical selective extinction function of \cite{Cardelli_1989_Ext}. We adopted a fixed ratio of extinction in V-band (=$5550\AA$), $A_V$, to color excess $E(B-V)$ of $R_V=3.1$, and adopted the galaxies' corresponding total extinction values $A_V$ from \cite{Schlafly_Finkbeiner_2011_deredden}. We selected spaxels based on an S/N>3 criterion for the $H\beta$, $[OIII]\lambda 5007$, $[NII]\lambda 6583$, and $H\alpha$ emission lines to ensure an accurate line detection. We chose this criterion as it corresponds to a certain flux limit that allows us to get accurate measurements. 
 
Only spaxels falling within the star-forming region of the $[NII]/H\alpha$ versus $[OIII]/H\beta$ Baldwin-Philips-Terlevich (BPT, \citealt{Baldwin_1981_BPT}) diagnostic diagram employing the empirical line calibration by \cite{Kauffmann_2003} (see Fig.~\ref{fig:BPT_diagrams}) were taken into account. We did not correct the empirical and theoretical line calibrations for redshift due to the only faint effect present at our target redshift of $z\sim 0.3$ or corrected for underlying stellar absorption, which could result in a slight overestimation of the $[OIII]/H\beta$ ratio. In total, 9825 spaxel fulfilled our S/N criterion, while 6299 were classified as SF-spaxel and utilized for the following spatially resolved analysis.

\section{Methods}\label{sec:methods}
\subsection{Emission line fits and stellar masses}\label{sec:fluexes_meth}

We determined emission line fluxes and stellar masses via the population spectral synthesis code Fitting Analysis using Differential evolution Optimization (FADO; \citealt{Gomoes_2017_FADO}) using the library of simple stellar population (SSP) spectra from \cite{Bruzual_Charlot_2003_SSP}. FADO already corrects fluxes for underlying stellar absorption. FADO only provides formal errors in all quantities, meaning that the errors are based on the goodness of the fits and the convergence of the different evolutionary threads, resulting in the best-fitting solution obtained via differential genetic optimization. These errors should be seen as low limits to the quantities' true limits, most specifically, for the stellar masses. As a consistency check, we also computed emission line fluxes via MPDAF \citep{Bacon_2016_MPDAF} and obtained measurements done via GIST \citep{Bittner_2019_GIST}, and found good agreement between the measurements yielded by the three different tools. 

We corrected all emission lines for dust extinction via the $H\alpha/H\beta$ Balmer decrement following the equations introduced by \cite{Calzetti_2001_extinction}, which assume a \cite{Cardelli_1989_Ext} extinction curve, assuming an intrinsic flux ratio of $H\alpha/H\beta = 2.86$ and $R_V = 3.1$. Lastly, we corrected surface densities for inclination by assuming an infinitely thin disk and a simplified inclination angle of $\cos(\Theta) = b^{\prime}/a^{\prime}$:
\begin{equation}\label{eq:inclination_correction}
    \Sigma_{corrected} = \frac{b^{\prime}}{a^{\prime}} \cdot \Sigma_{observed}.
\end{equation}
We note that $a^{\prime}$ and $b^{\prime}$ refer to the major and minor axes.

\subsection{Star formation rate surface densities}\label{sec:methods_SFR}

The star formation rates were calculated via the extinction-corrected $H\alpha$ luminosity using the SFR calibration from \cite{Kennicutt_1998_SFR} assuming a Salpeter IMF:
\begin{equation}
\text{SFR}(M_{\odot} \ yr^{-1}) = 7.9 \cdot 10^{-42} \  L_{H\alpha}(erg \ s^{-1}).
\end{equation}

We converted SFRs from a Salpeter IMF to Chabrier IMF via the conversion factor from \cite{Driver_2013_IMF}: $M_{\text{Chabrier}} = M_{\text{Salpeter}}/1.53$. We converted spatially resolved star formation rates into surface densities by dividing each spaxel's SFR by its area: $\Sigma_{SFR}=SFR/A \ (M_{\odot} yr^{-1} kpc^{-2})$. We record an average error of $0.04$ dex for $\log(\Sigma_{SFR})$.

\subsection{Gas-phase metallicities}\label{sec:metals}

We utilized an optical strong-line calibration to measure gas-phase metallicities. We used the $O3N2$ index, which is defined as 
\begin{equation}
    O3N2 = \log{\left( \frac{[OIII]\lambda 5007/H\beta}{[NII]\lambda 6583/H\alpha}  \right)}
\end{equation}
and the calibration by \cite{Marino_2013_metallicity} with O3N2 ranging from -1.1 to 1.7, corresponding to a range in metallicity of $8.17 \lesssim 12+\log(O/H) \lesssim 8.77$:
\begin{equation}
    12+\log{(O/H)} = 8.533-0.214 \cdot O3N2.
\end{equation}

The average error for the metallicity calibration derived via oxygen with this O3N2 diagnostic is $\sim 0.08$ dex. For our sample, 39 spaxels fall outside of the metallicity range defined by \cite{Marino_2013_metallicity}, and we henceforth excluded them from our analysis.

\section{Results}\label{sec:results}

In the following sections, we present several results to explore the SFMS, MZR, and FMR in the MAGPI sample. We compare these results to previous works. Differences in IMFs between our work and others have been adjusted. All emission lines used in this analysis are corrected for intrinsic extinction based on the Balmer decrement as described in section~\ref{sec:fluexes_meth}, and we selected only spaxels classified as star-forming in the BPT diagram by \cite{Baldwin_1981_BPT} using the \cite{Kauffmann_2003} empirical separation line.

\subsection{Ionization sources}

\subsubsection{Diagnostic diagrams}

\begin{figure*}
    \centering
    
    \begin{subfigure}{0.3\textwidth}
        \includegraphics[width=\linewidth]{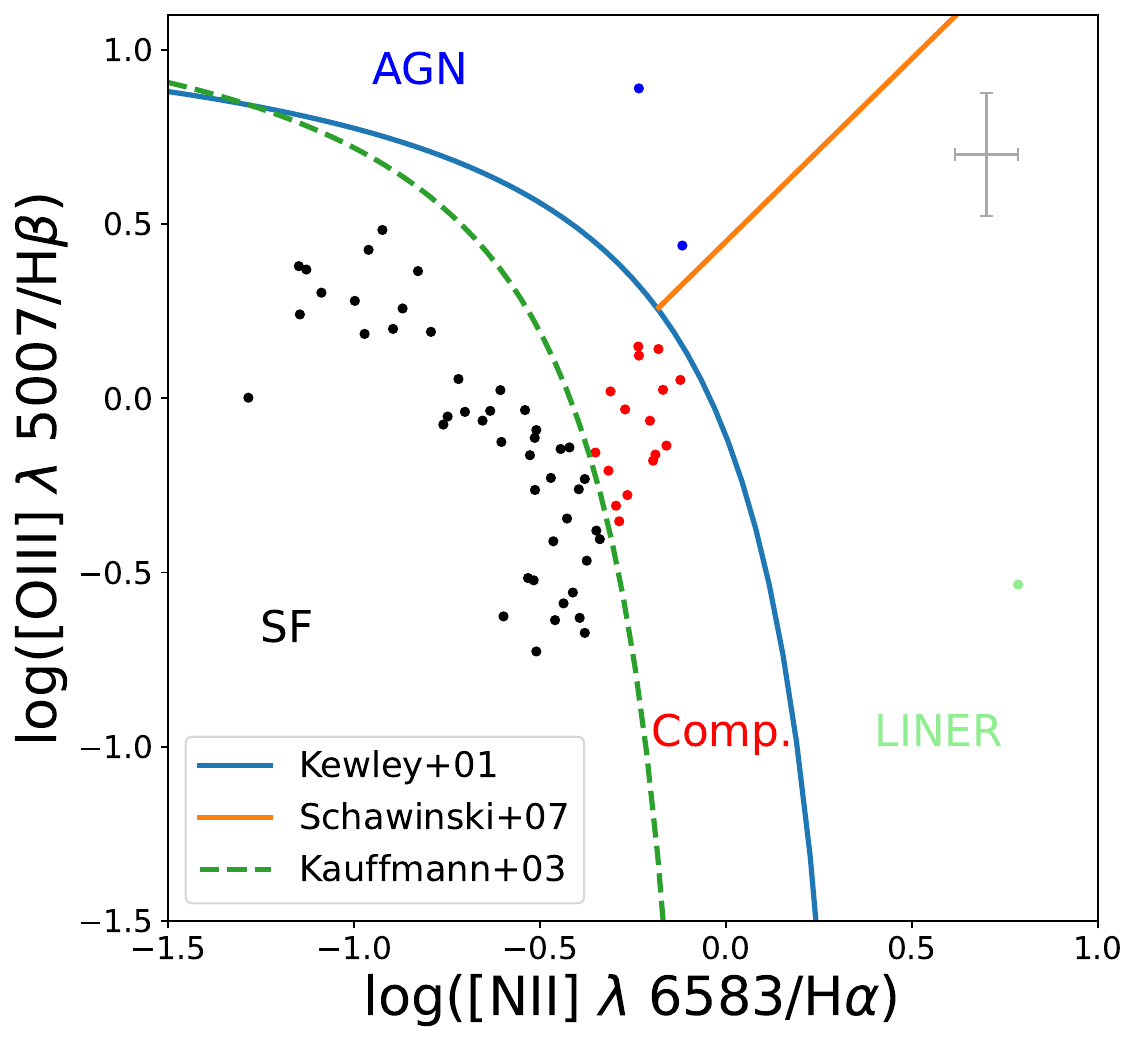}
    \end{subfigure}
    \hfill
    \begin{subfigure}{0.3\textwidth}
        \includegraphics[width=\linewidth]{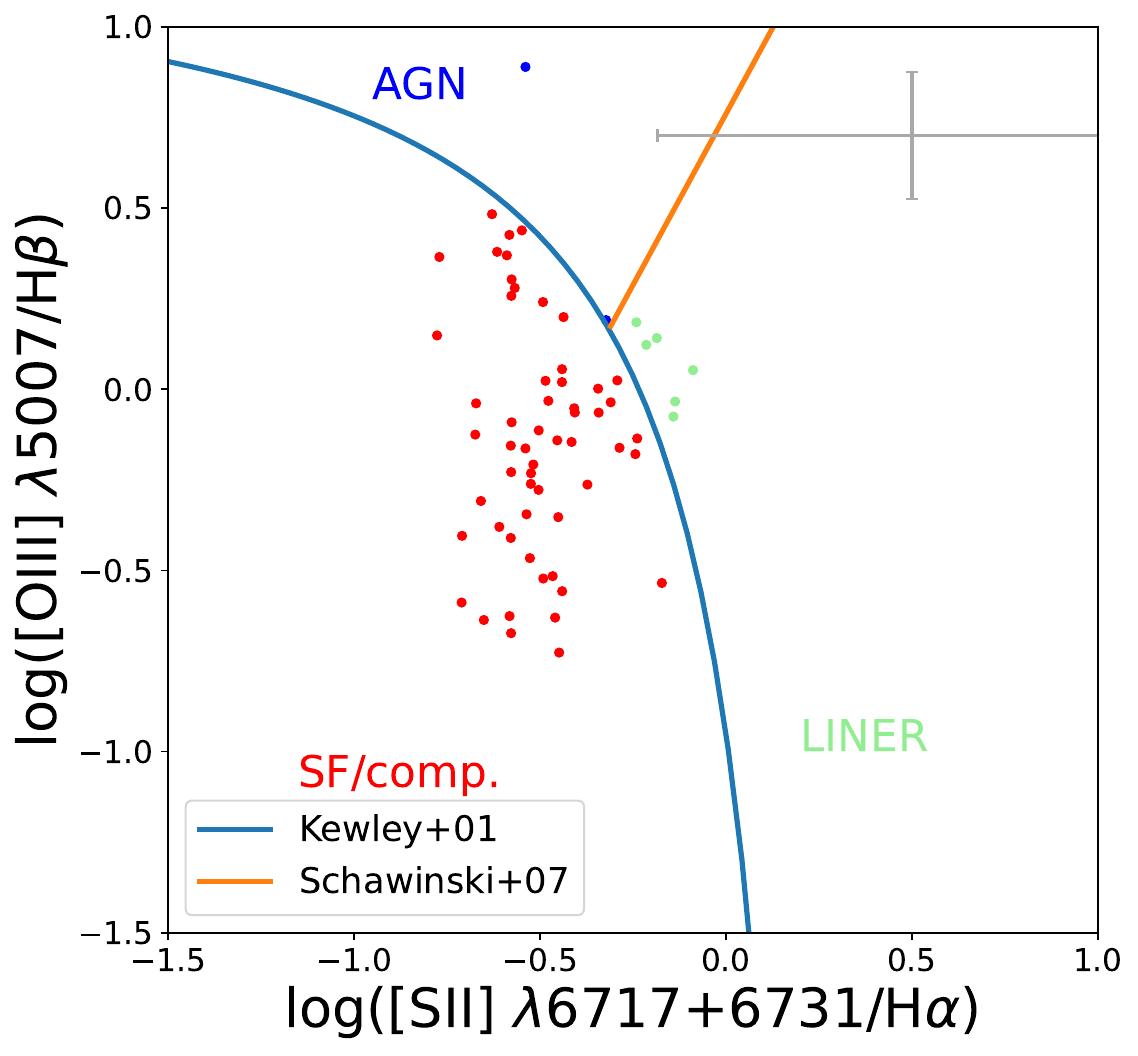}
    \end{subfigure}
    \hfill
    \begin{subfigure}{0.3\textwidth}
        \includegraphics[width=\linewidth]{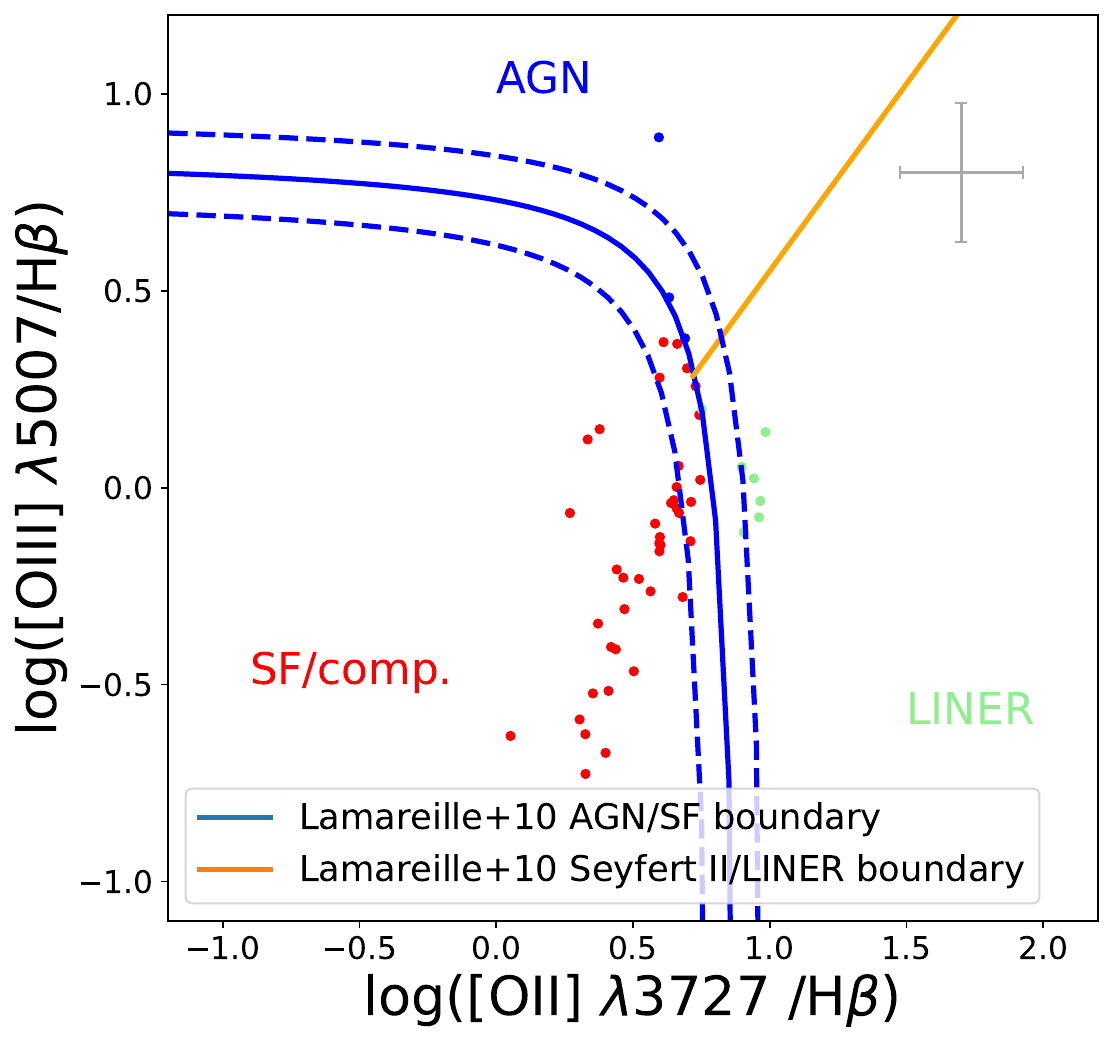}
    \end{subfigure}
    
    \vspace{0.5cm}

     \begin{subfigure}{0.3\textwidth}
        \includegraphics[width=\linewidth]{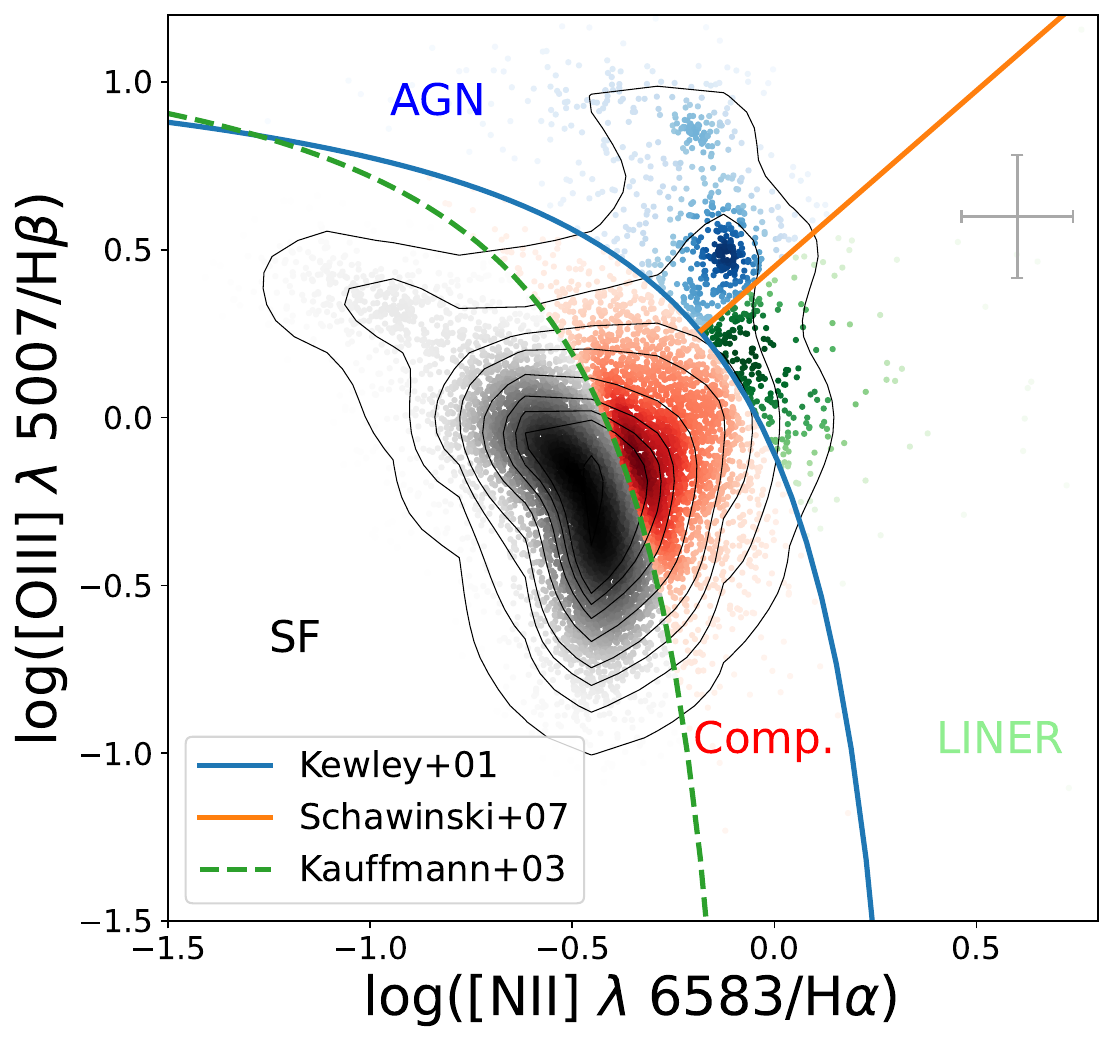}
    \end{subfigure}
    \hfill
    \begin{subfigure}{0.3\textwidth}
        \includegraphics[width=\linewidth]{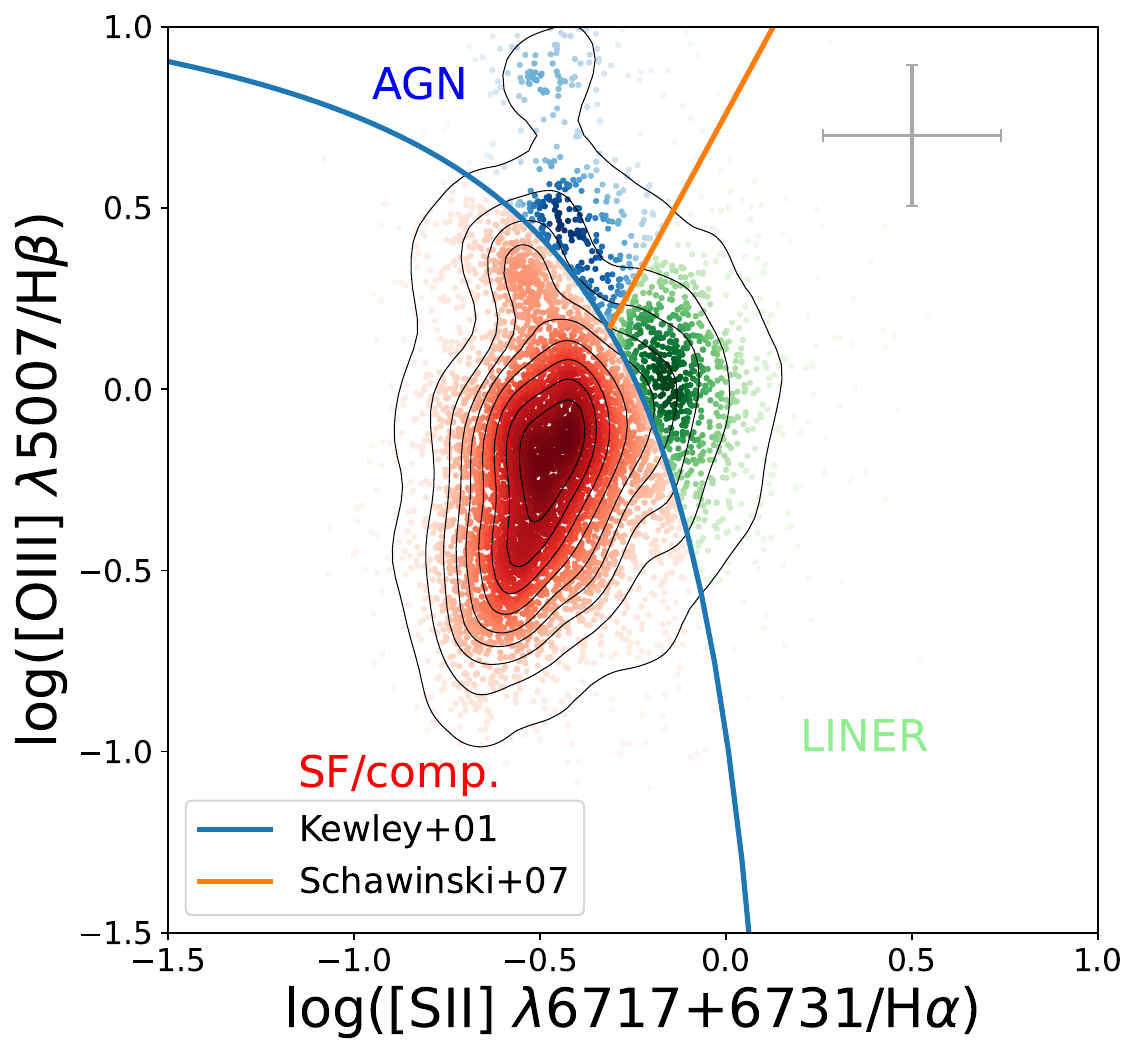}
    \end{subfigure}
    \hfill
     \begin{subfigure}{0.3\textwidth}
        \includegraphics[width=\linewidth]{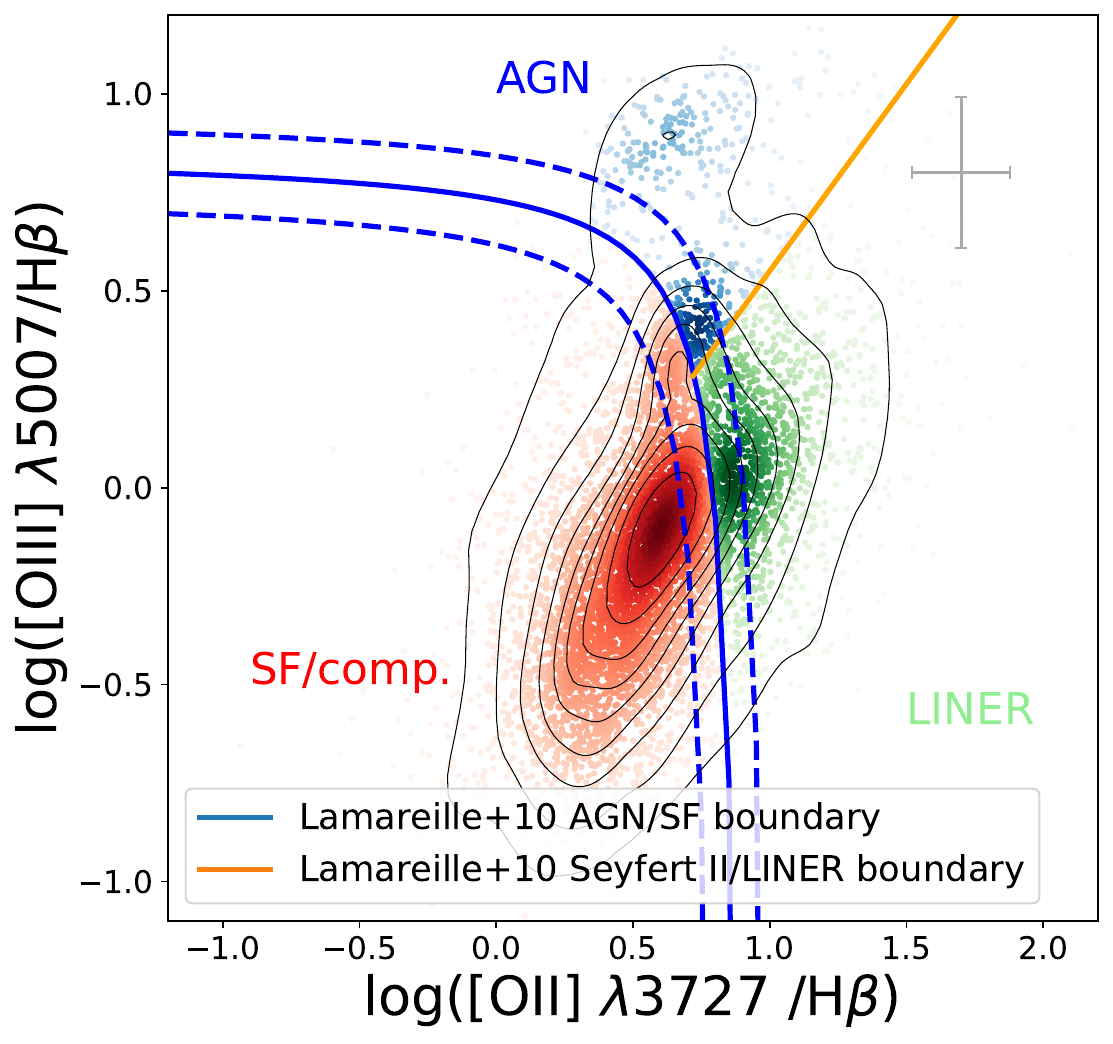}
    \end{subfigure}

    \caption{Global and spatially resolved diagnostic diagrams. \textit{Top:} Diagnostic diagrams by \cite{Baldwin_1981_BPT} (left), \cite{Veilleux_1987_SIIBPT} (middle), and \cite{Lamareille_2010_OIIBPT} (right) for the galaxies' integrated flux measurements. The gray error bars in the top right indicate all diagrams' mean flux ratio errors along the x- and y-axis. In the case of the \cite{Lamareille_2010_OIIBPT} diagram, only galaxies with $[OII] \lambda 3727$ measurements via FADO were included.  \textit{Bottom:} The same diagnostic diagram of all spaxels satisfying the $S/N>3$ criteria in all four emission lines - $[OIII] 5006 \AA$, $H\beta$, $H\alpha$, and $[NII] 6585 \AA$. The diagnostic diagram by \cite{Baldwin_1981_BPT} distinguishes between star-forming (black), composite (red), AGN (blue), and LINER (green) emission while the diagrams by \cite{Lamareille_2010_OIIBPT} and \cite{Veilleux_1987_SIIBPT} both only differentiate between star-forming and possibly composite (red), AGN (blue), and LINER (green).} 
    \label{fig:BPT_diagrams}
\end{figure*}

The different ionization sources present in our sample of galaxies were investigated via three diagnostic diagrams, which use a set of four strong emission lines. By applying these diagnostics to our spatially resolved emission line flux measurements, we can reliably distinguish between star-forming (SF), Seyfert (AGN), low ionization nuclear emission-line regions (LINER), and composite galaxies or spaxels. Composite galaxies or spaxels are regions where photoionization is powered by star formation, an AGN, and possibly other mechanisms responsible for LINER emissions, such as shocks or cosmic rays. We utilized the classical BPT diagram by \cite{Baldwin_1981_BPT} which makes use of the $[OIII]/H\beta$ and $[NII]/H\alpha$ emission line ratios, the $[OIII]/H\beta$ versus $[SII]/H\alpha$ diagnostic diagram of \cite{Veilleux_1987_SIIBPT}, as well as the $[OIII]/H\beta$  versus $[OII]/H\beta$ diagram of \cite{Lamareille_2010_OIIBPT}.

Figure~\ref{fig:BPT_diagrams} shows the results for the classical BPT diagram on the left-hand side: the results on the top are for integrated galaxy values, while the results shown at the bottom are for a spaxel-by-spaxel analysis. The differently colored lines within the diagram act as a separation between the different ionizing mechanisms. The solid blue curve of \cite{Kewley_2001} depicts the theoretical line that separates SF and composite galaxies or spaxels from AGN and LINER galaxies or spaxels. In contrast, the dashed green line of \cite{Kauffmann_2003} is an empirical calibration that further distinguishes between SF and composite galaxies or spaxels. Additionally, the solid orange curve of \cite{Schawinski_2007} represents the separation line between AGN and LINER galaxies or spaxels. Our results show a sample dominated by star-forming ($64\%$) and composite ($27\%$) spaxels, and the remaining $9\%$ spaxels are classified as AGN or LINER. 

The middle panel of Fig.~\ref{fig:BPT_diagrams} depicts the $[OIII]/H\beta$ versus $[SII]/H\alpha$ diagnostic diagram of \cite{Veilleux_1987_SIIBPT}. Here, the solid blue curve represents the theoretical separation curve of \cite{Kewley_2001}, which separates SF galaxies or spaxels from AGNs and LINERS. In contrast, the solid orange curve represents the separation line of \cite{Schawinski_2007}, which helps distinguish between AGN and LINER galaxies or spaxels. Again, our sample is shown to be dominated by spaxels ionized by star formation, with roughly $82\%$ classified as such. Compared to the $3\%$ of LINER emission classified by the BPT diagram, $12\%$ of all spaxels amount to LINER emission in the \cite{Veilleux_1987_SIIBPT} diagram. 

Lastly, the right panel of Fig.~\ref{fig:BPT_diagrams} depicts the \cite{Lamareille_2010_OIIBPT} diagnostic diagram, which is specifically designed to use emission lines from the bluer side of galaxies spectra to avoid them being redshifted out of the wavelength range of optical spectrographs. It can, therefore, distinguish between star-forming and AGN galaxies or spaxels at intermediate redshifts (z>0.3) where the danger of the $[NII]$, $[SII]$, and $H\alpha$ emission lines not being observable is present. The solid blue line separates AGN from star-forming galaxies or spaxels. Its $0.1$ dex uncertainty regions are also shown as dashed blue lines. Furthermore, an empirical line to distinguish between LINER and AGN galaxies or spaxels is also introduced. Both of these lines were taken from \cite{Lamareille_2010_OIIBPT}. Again, our sample is observed to be dominated by SF spaxels. 

Within the global BPT diagram shown in the top left of Fig.~\ref{fig:BPT_diagrams}, $70\%$ of galaxies are classified as SF. This percentage increases to $80\%$ for both the \cite{Veilleux_1987_SIIBPT} diagram (top middle) and \cite{Lamareille_2010_OIIBPT} diagram (top right). To conclude, in both the spaxel-by-spaxel and global analyses via integrated galaxy fluxes, we confirm that SF and composite galaxies dominate our sample.

\subsubsection{Spectral decomposition method}\label{sec:SDM}

We also applied the spectral decomposition method from \cite{Davies16, Davies17} to gain deeper insights into the ionizing sources dominating our sample. This method lets us dissect and filter out the contributions from star formation, AGN activity, and LINER emission to the luminosity of selected emission lines of individual spaxels. The method is based on the diagnostic diagram by \cite{Baldwin_1981_BPT}. Here, we first have to choose three basis spectra, each representing one of the three ionization sources: star formation, AGN emission, and LINER emission. In our case, we opted not to select specific spaxels as base spectra, as we analyzed our entire sample simultaneously and did not want the fluxes from one specific galaxy to have such a significant impact on the outcome. Instead, we elected to compute the median fluxes of the four emission lines $[OIII]$, $H\beta$, $[NII]$, and $H\alpha$ in each BPT region representing the three ionizing mechanisms. For instance, we can compute a star formation base spectrum via the line ratios computed from the median fluxes taken over all spaxels in the star-forming region of the BPT diagram. To check whether our chosen median base spectra accurately represent each ionization mechanism, we also computed and plotted them within the diagnostic diagram from \cite{Veilleux_1987_SIIBPT} and found good agreement with their respective location within the diagram's different sections and their location within the BPT diagram. Furthermore, the second step in this decomposition method is to apply the following linear superposition equation defined by \cite{Davies17}:

\begin{equation}\label{eq:SDM}
    L_i(j) = m(j) \cdot L_i(HII) + n(j) \cdot L_i(AGN) + k(j) \cdot L_i(LINER),
\end{equation}
where $L_i(j)$ refers to the luminosity of any emission line $i$ of any spaxel $j$. $L_i(HII)$, $L_i(AGN)$, and $L_i(LINER)$ are the luminosities for any emission line $i$ of the star-forming, AGN, and LINER base spectra. $m$ is the superposition coefficient for star formation, $n$ for AGN emission, and $k$ for LINER emission. These three coefficients vary between spectra but are the same for all emission lines within a spectrum. In summary, this equation represents the linear superposition of the line luminosities of the different ionizing mechanisms that make up every emission line of every spaxel. We computed the superposition coefficients by imposing a least-square minimization on Eq.~\ref{eq:SDM} via the Levenberg-Marquardt nonlinear fitting routine from lmfit \citep{Newville_2016_LMFIT}. The luminosities in Eq.~\ref{eq:SDM} represent the extinction-corrected luminosities of the following four emission lines: $[OIII]\lambda5007$, $H\alpha$, $[NII]\lambda 6584$, and $[SII]\lambda 6718,6731$. Afterward, we utilized the superposition coefficients to calculate the luminosities of the emission lines of interest associated with star formation, AGN activity, and LINER emission by simply multiplying each superposition coefficient with its respective basis spectrum and the $H_\alpha$ luminosity. We multiply with $L_{H\alpha}$ as in an earlier step, we normalize both the luminosities $L_i(j)$ on the left-hand side of Eq.~\ref{eq:SDM} and the luminosities of the basis spectra to an $H\alpha$ luminosity of 1. 

We note that the resulting luminosities can be used for computing SFRs, which is discussed in section~\ref{sec:rSFMS}. However, since all spaxels are scaled versions of the star-forming regions basis spectrum, it is not possible to compute flux ratios, ergo gas-phase metallicities, as all spaxels would have very similar or the same value. Additionally, potential limitations of the spectral decomposition method stem from the dependence of emission line luminosities on various factors, including the mix of ionization mechanisms and the metallicity and ionization parameter of the gas \citep{Davies17}. On the contrary, this method is advantageous as it enables the inclusion of more spaxels than the conventional method of only SF spaxels. Our analysis encompassed 9475 spaxels, a substantial increase over the 6299 spaxels included via the traditional method.

\begin{figure*}
    \centering
    
    \begin{subfigure}{0.3\textwidth}
        \includegraphics[width=\linewidth]{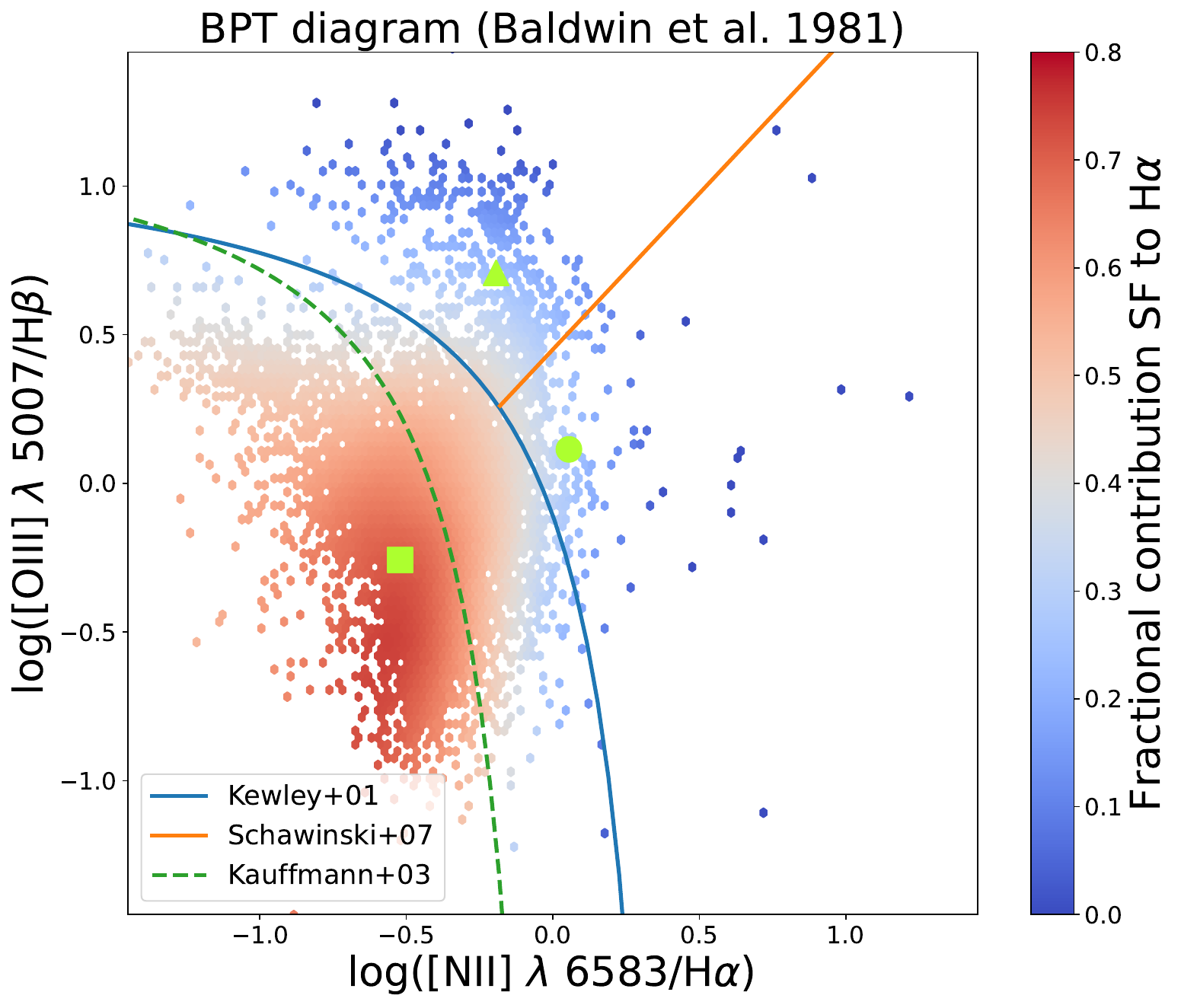}
        \label{fig:sub1}
    \end{subfigure}
    \hfill
    \begin{subfigure}{0.3\textwidth}
        \includegraphics[width=\linewidth]{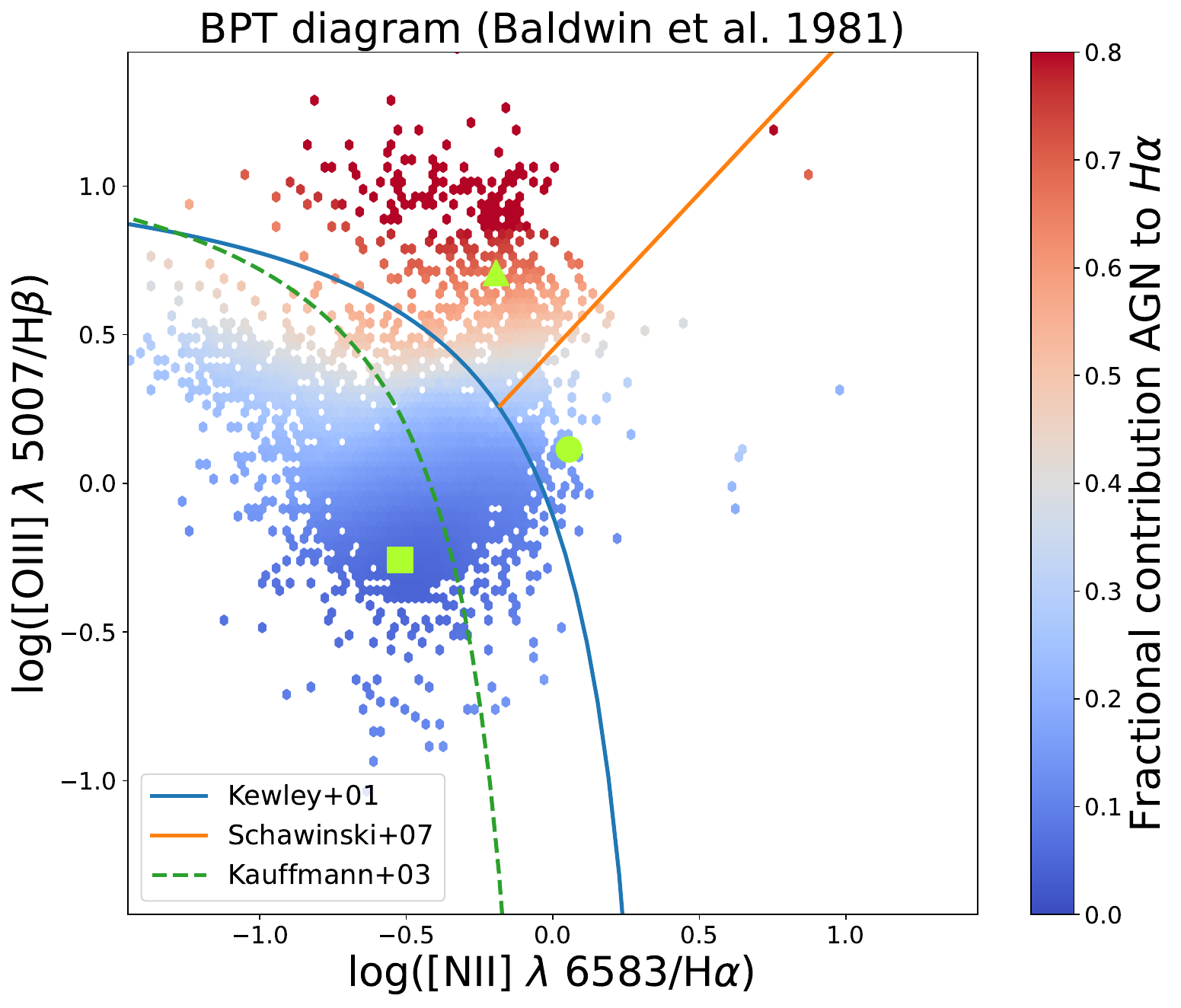}
        \label{fig:sub3}
    \end{subfigure}
    \hfill
    \begin{subfigure}{0.3\textwidth}
        \includegraphics[width=\linewidth]{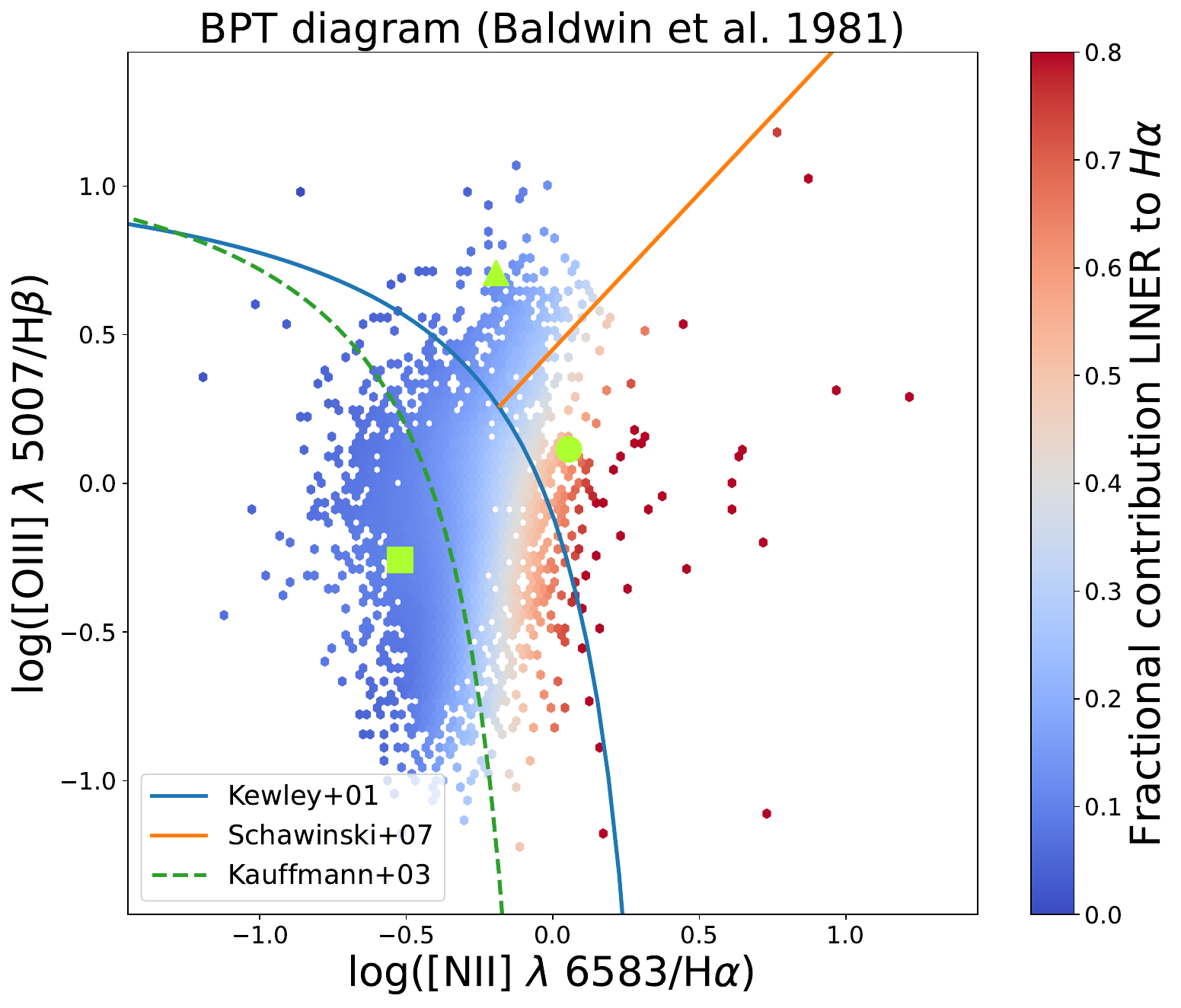}
        \label{fig:sub3}
    \end{subfigure}

    \vspace{0.5cm} 

    \begin{subfigure}{0.3\textwidth}
        \includegraphics[width=\linewidth]{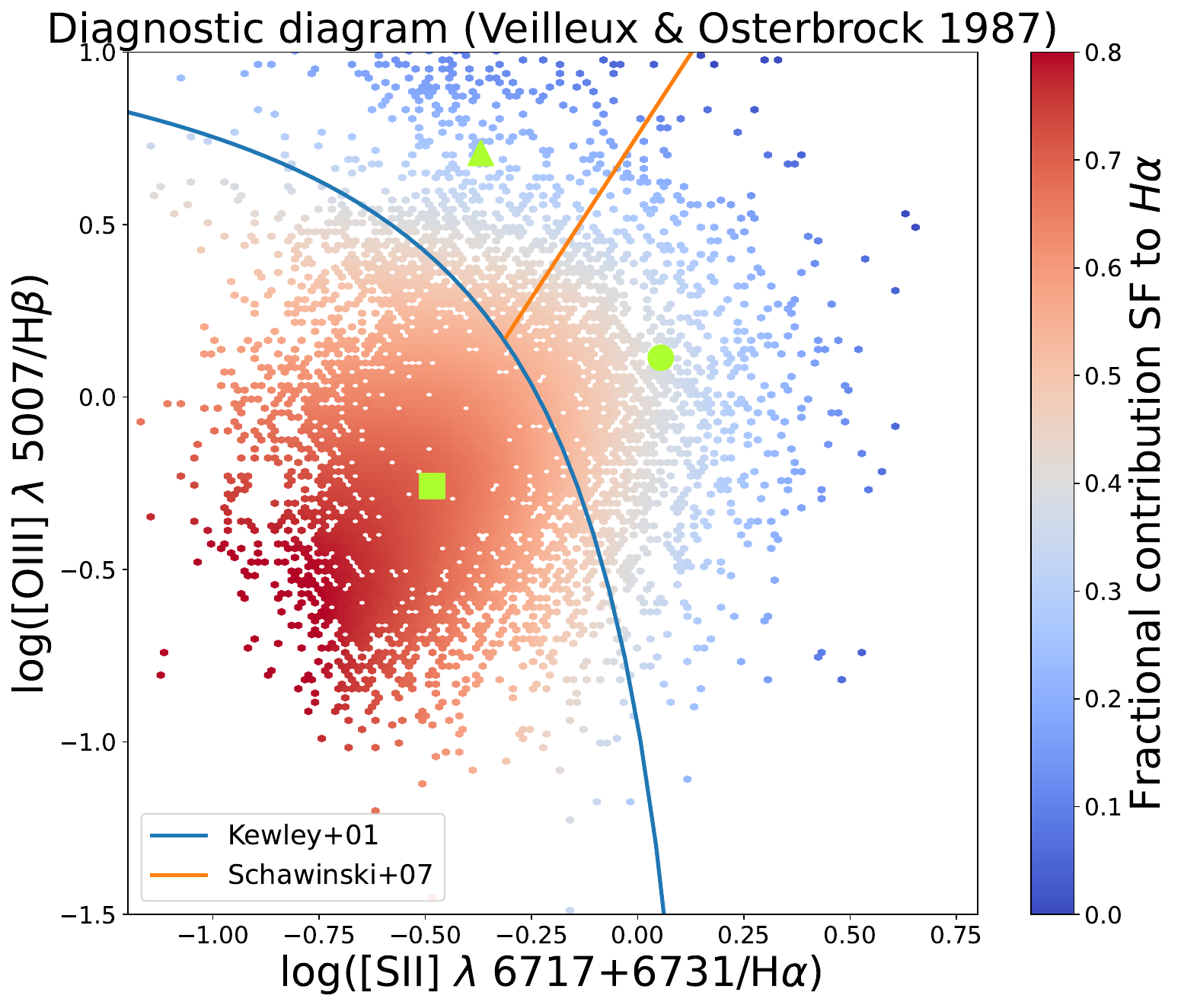}
        \label{fig:sub4}
    \end{subfigure}
    \hfill
    \begin{subfigure}{0.3\textwidth}
        \includegraphics[width=\linewidth]{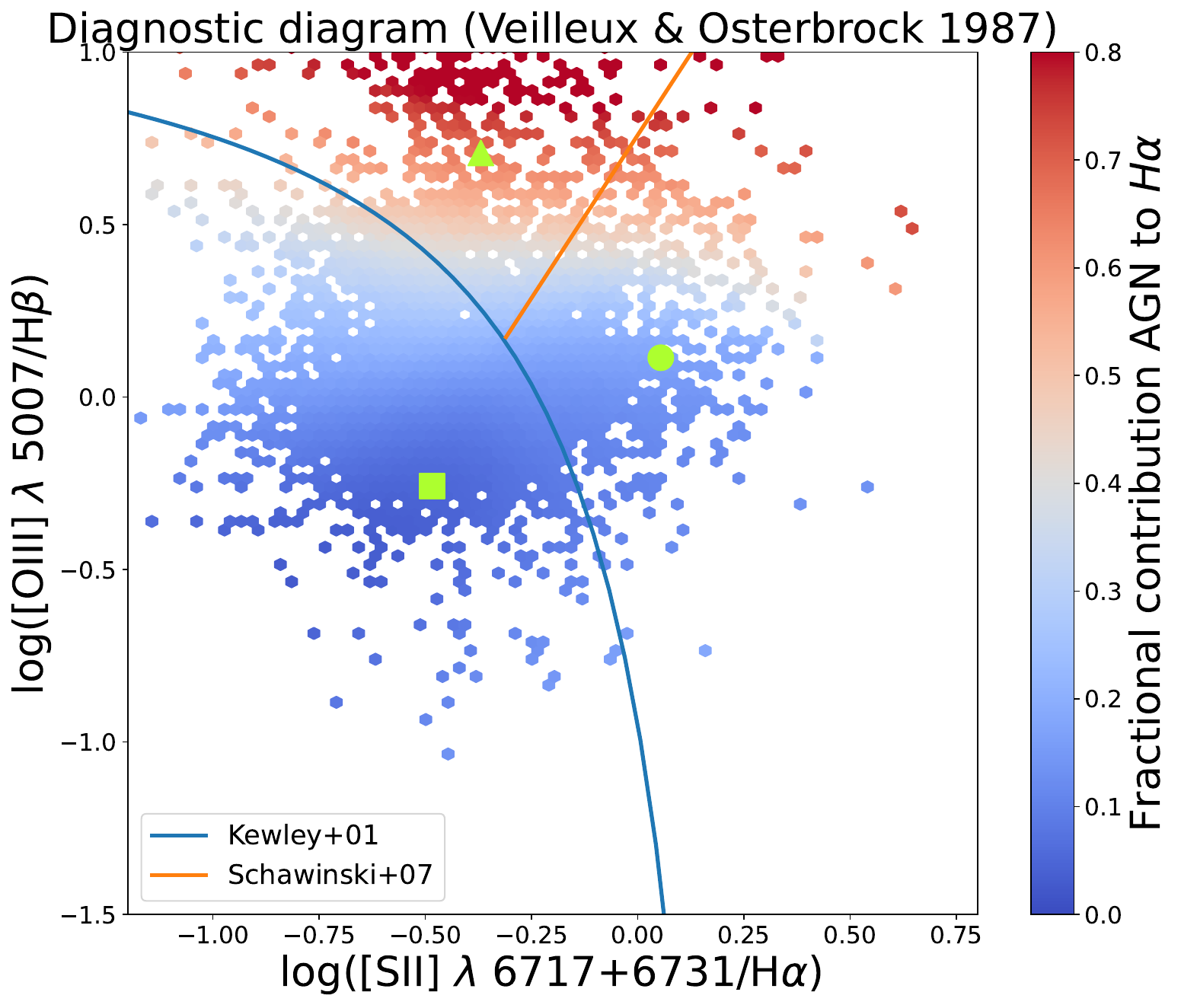}
        \label{fig:sub6}
    \end{subfigure}
    \hfill
    \begin{subfigure}{0.3\textwidth}
        \includegraphics[width=\linewidth]{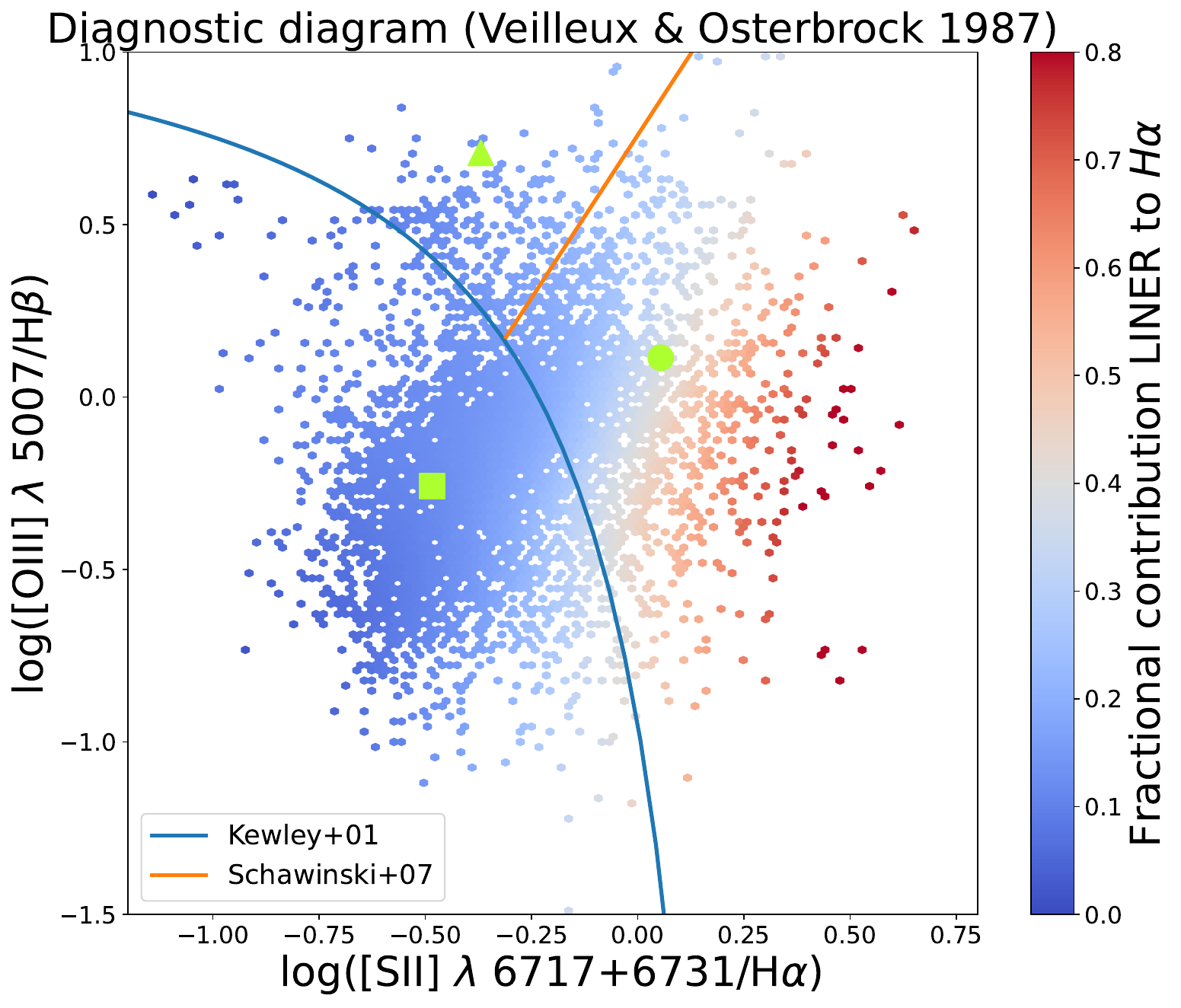}
        \label{fig:sub6}
    \end{subfigure}

    \caption{Diagnostic diagrams by \cite{Baldwin_1981_BPT} (top row) and \cite{Veilleux_1987_SIIBPT} (bottom row) with emission line flux ratios extracted from the total emission line fluxes of the individual spaxels. The lime green icons indicate the base spectra for star formation (square), AGN emission (circle), and LINER emission (triangle). Our data points shown here are color-coded using the fraction of $H\alpha$ emission that is ascribable to SF (left), AGN emission (middle), and LINER emission (right). The data points are smoothed over via LOESS \citep{Cappellari2013b_Loess}.}
    \label{fig:BPT_diagrams_SDM}
\end{figure*}

Examining the effectiveness of the spectral decomposition approach from \cite{Davies17} involves assessing the consistency in emission fractions attributed to individual ionization mechanisms across line ratios. This consistency should ideally peak at the line ratios corresponding to the relevant basis spectrum and gradually decline as the line ratios approach those of another basis spectrum. In this sense, we should be able to observe that the different empirical and theoretical lines used to distinguish between the ionization mechanisms match with a decrease in the fractional contributions. Results are shown in Fig.~\ref{fig:BPT_diagrams_SDM} where the upper three panels present the BPT diagram by \cite{Baldwin_1981_BPT} and the lower three panels show the diagnostic diagram by \cite{Veilleux_1987_SIIBPT}. The data points are color-coded according to their fractional contribution of SF (left), AGN (middle), and LINER (right) to the $H\alpha$ emission line. 

The line calibration by \cite{Kauffmann_2003} aligns quite well with the fractional contribution of SF to $H\alpha$. Most notably, the contribution of AGN is strong throughout the entire AGN region of both diagnostic diagrams and extends deep into the LINER region. Lastly, similar to the SF region, the spaxels with high contribution from LINER emission are also well contained via the \cite{Kewley_2001} and \cite{Schawinski_2007} lines. 

It is worth mentioning that in the case of the contributions from AGN and LINER, we had to omit a significant amount, approximately a third of all spaxels, from Fig.~\ref{fig:BPT_diagrams_SDM} for which this method could not produce meaningful values as the resulting fractional contributions were too close to zero. These spaxels lie within the parts of the diagrams that are furthest away from the respective emission. For example, in the top right panel showing the fractional contribution from LINER emission to the $H\alpha$ emission line, some spaxels of the AGN and SF regions were omitted. 

We conclude from the spectral decomposition method that our sample is predominately dominated by SF with additional contributions from LINER and AGN. The average fractional contribution of star formation to the $H\alpha$ emission line accounts for $\sim 50-60\%$ of
the ionized gas emission, whereas both AGN and LINER emission account for roughly $\sim 20 \%$ each.

\subsection{Global relations}

This section briefly summarizes the resulting global versions of the SFMS and MZR relations. We implemented two approaches to obtain global values for the stellar mass $M_{*}$ and SFR. First, we used the integrated spectrum of each galaxy and processed it through FADO as described in section~\ref{sec:fluexes_meth}, which gives us a total stellar mass estimate as well as a global $H\alpha$ flux measurement, which is then used to compute a global SFR for each galaxy as described in section~\ref{sec:methods_SFR}. Secondly, we computed the sum of all resolved spaxel values that fulfill our S/N criteria within each galaxy to obtain measurements for global properties. 

Generally, SFRs obtained by totaling the resolved values are expected to be smaller than integrated SFRs. As with the former method, some spaxels are excluded from the calculations, resulting in a weaker overall flux, ultimately underestimating the global SFR. Indeed, we observe an average difference of $2.2 \ M_{\odot} yr^{-1}$ between integrated and total resolved global SFRs. Additionally, integrated and total resolved stellar masses obtained via FADO were compared to those obtained via ProSpect by MAGPI team members. We find that ProSpect masses are, on average, $0.04$ dex higher than the integrated FADO masses and $0.14$ dex higher than the total resolved FADO masses. In the following sections concerning the spatially resolved results, we applied global integrated measurements whenever global parameters were used for the stellar mass $M_{*}$ and SFR. 

\begin{figure}
    \centering
    \begin{subfigure}{0.45\textwidth}
        \includegraphics[width=\linewidth]{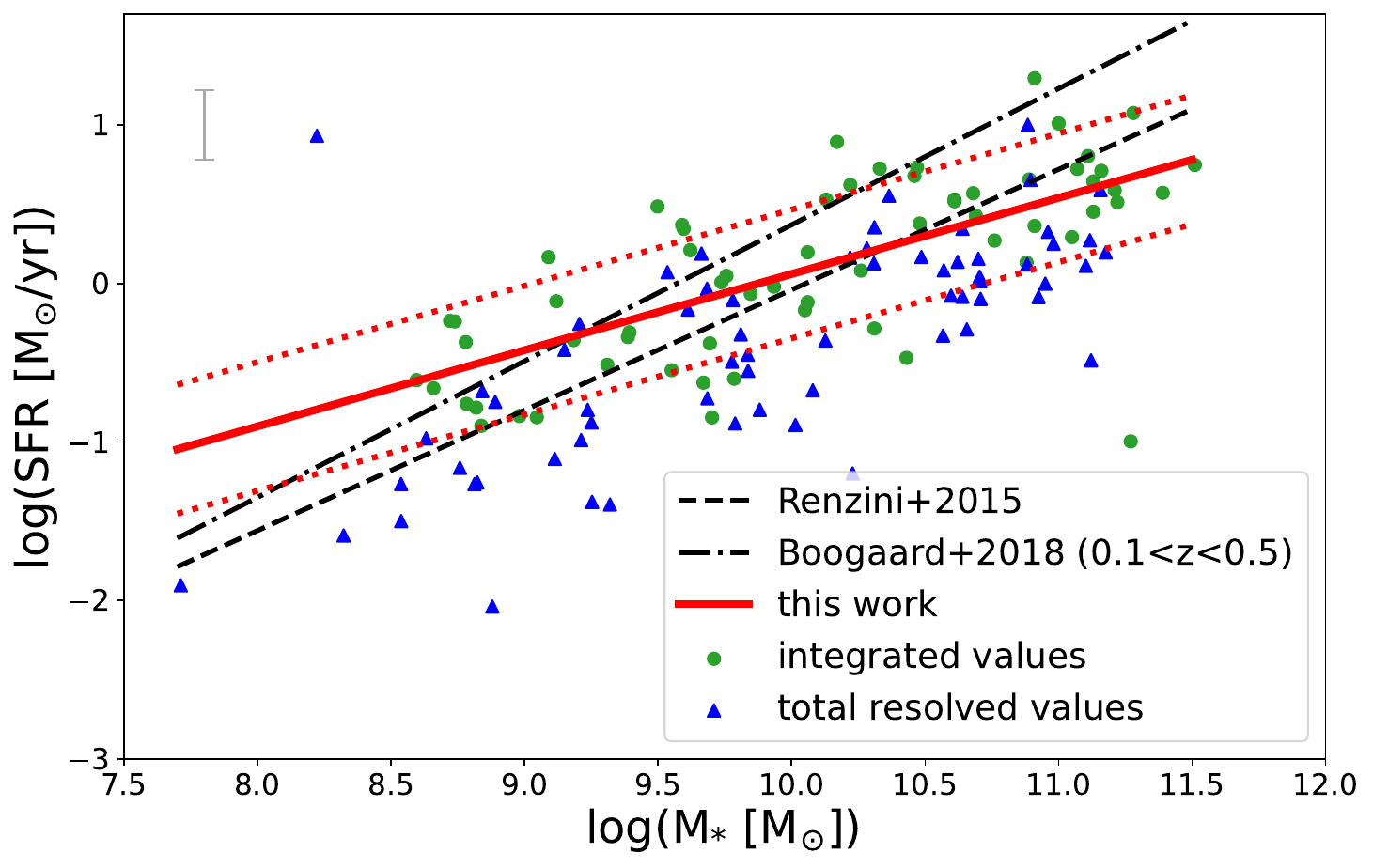}
    \end{subfigure}

    \vspace{0.5cm} 

    \begin{subfigure}{0.45\textwidth}
        \includegraphics[width=\linewidth]{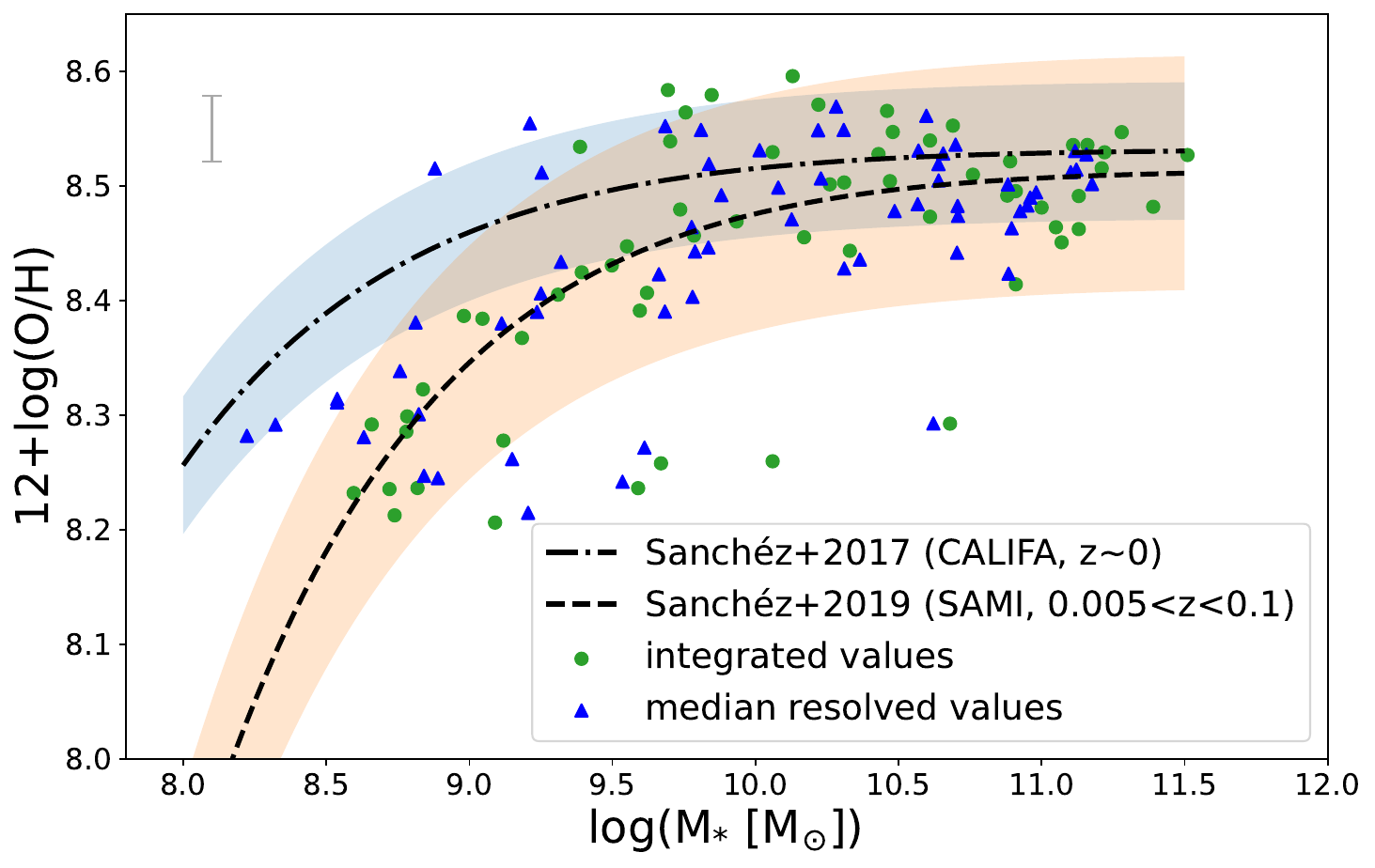}
    \end{subfigure}
    \caption{Global relations for our sample. \textit{Top:} Global SFMS. The green circles depict the integrated measurements for SFR and $M_{*}$, while the blue triangles represent the sum of all the resolved spaxel values within each galaxy. The dashed black line represents the fit from \cite{Renzini_2015_SFMS}. The dash-dotted black line shows the fit from \cite{Boogaard_2018_SFMS}. Our own linear fit results are shown in red with its $\sim 0.4$ dex scatter as a dotted red line. \textit{Bottom:} Global MZR. The integrated $12+\log(O/H)$ as green circles are calculated using the integrated flux measurements, while the blue triangles represent the median resolved gas-phase metallicity of each galaxy. The dashed black line and $0.102$ dex scatter in orange represents the MZR fit from \cite{Sanchez_2019_MZR_SAMI} while the dotted black line and $0.06$ dex scatter in blue stems from \cite{Sanchez_2017_MZR_global}.}
    \label{fig:global_relations}
\end{figure}

\subsubsection{Star formation main sequence}\label{sec:SFMS}

The top panel of Fig.~\ref{fig:global_relations} depicts the global SFMS for both the integrated values (green circles) and total resolved values (blue triangles). Furthermore, we also utilized two different fits for comparisons: the function by \cite{Renzini_2015_SFMS}, which evaluated the SFMS based on SDSS data, and by \cite{Boogaard_2018_SFMS}, which utilized SFRs derived from $H\alpha$ and $H\beta$ (only for galaxies with $z>0.42$) and is defined for a redshift range of $0.1<z<0.5$. Both calibrations were derived via integrated measurements and also best align with our integrated values. Additionally, a subgroup of galaxies exhibits relatively low SFRs according to their total resolved values, especially when compared to their integrated measurements. A possible explanation could be that many of their spaxels did not pass our S/N criteria within these galaxies, resulting in their being omitted when computing global parameters. 

We applied an ordinary linear least-square fitting and utilized the following function in log-log space considering individual errors: 
\begin{equation}\label{eq:SFMS}
    \log(SFR) = b \cdot  \log (M_{*}/M_{\odot}) + a.
\end{equation}

This results in the following fit: $b = 0.481 \pm 0.242$ and $a = -4.749 \pm 2.412$. We measure a $1\sigma$ scatter, derived by calculating the standard deviation of the residual between the observed and fitted $\log(SFR)$, of $\sim 0.4$ dex.

\subsubsection{Mass metallicity relation}\label{sec:MZR}

The bottom panel of Fig.~\ref{fig:global_relations} shows the global gas MZR. Corresponding to the resolved stellar mass measurements, gas-phase metallicities of each galaxy are their median metallicity over all spaxel values. Integrated gas-phase metallicities were computed by taking the integrated flux measurements. In this case, no overall significant difference in the distribution of the data points between median and integrated values is apparent, most likely due to shallow metallicity gradients within this sample where removing low S/N spaxels does not alter the overall line ratios. We compare our results to those found by \cite{Sanchez_2017_MZR_global}, who investigated the MZR via CALIFA data at $z\sim 0$, and by \cite{Sanchez_2019_MZR_SAMI}, who investigated the MZR via SAMI data at $0.005<z<0.1$. Both utilized the same $12+\log(O/H)$ calibration as us from \cite{Marino_2013_metallicity}. Our gas-phase metallicities for the MAGPI galaxies are overall quite high, with most of them situated around the MZR from CALIFA for the local Universe. Additionally, the relation by \cite{Sanchez_2019_MZR_SAMI}, which is closer to our results redshift-wise, aligns well with our gas-phase metallicity measurements. 

\cite{Maier_2015_MZR} investigated the MZR for massive galaxies at $0.5<z<0.75$ from the zCOSMOS survey \citep{Lilly_2007_zCOSMOS}. Taking global values on the MZR, observations show a very low offset in gas-phase metallicities and the resulting MZR between z=0 and those of \cite{Maier_2015_MZR} at $z \sim 0.5-0.75$. This aligns with our result of the MZR at $z \sim 0.3$ which is also located close to the local relation.

\subsection{Resolved SFMS}\label{sec:rSFMS}

\begin{figure}
    \centering
    
    \includegraphics[width=\linewidth]{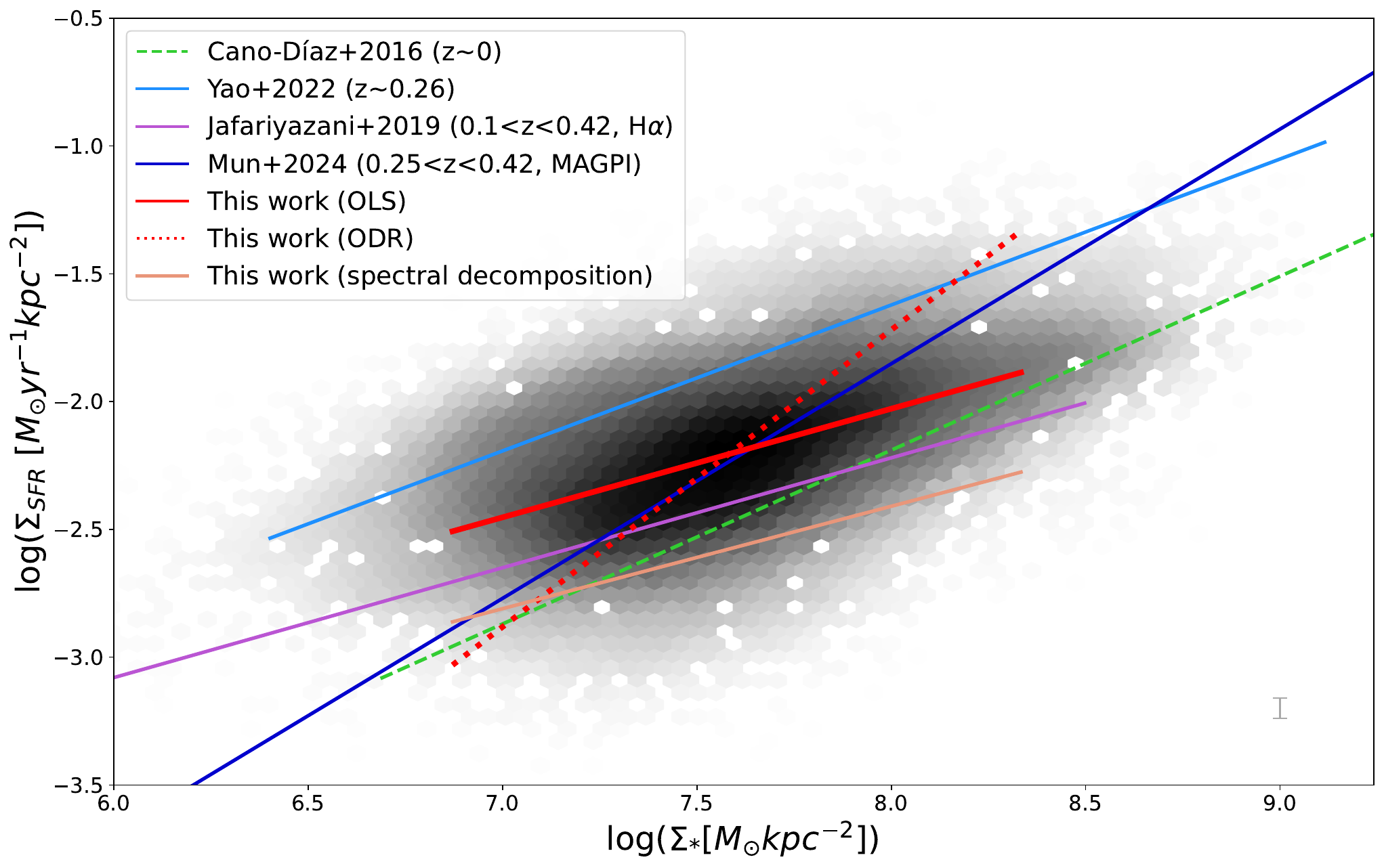}
    \caption{Results of the rSFMS compared to other works. The gray hexagonal bins depict the distribution of our data points, estimated via Gaussian kernels, where darker bins correspond to a higher density of data points. The solid red line represents our results using an ordinary least square (OLS) fitting and the dotted red line is for an orthogonal distance regression (ODR). The solid blue line is the result of \cite{Yao_2022} at $z\sim 0.26$, and the solid purple line is of \cite{Jafariyazani_2019_rSFMS} at $0.1<z<0.42$, both of which cover a similar redshift range to ours. The dashed green line comes from \cite{Cano_Diaz_2016_rSFMS} for the local Universe. As a comparison, we also present our results via OLS when applying the spectral decomposition method from \cite{Davies17} (see section~\ref{sec:SDM}) as a pink line. The gray errorbar shows the average uncertainties in $\Sigma_{SFR}$ and $\Sigma_{*}$. FADO only gives errors based on the goodness of the fit, stellar mass errors are very small, and the average $\Sigma_{*}$ is not visible within this plot.}
    \label{fig:rSFMS}
\end{figure}

Figure~\ref{fig:rSFMS} shows our results for the rSFMS. Within this and the following sections presenting our spatially resolved results, we utilized the previously selected SF-spaxel from the diagnostic diagram by \cite{Baldwin_1981_BPT}. The average $\log(\Sigma_{SFR})$ uncertainties are 0.04 dex. We applied both an OLS and orthogonal distance regression (ODR) fitting via Eq.~\ref{eq:SFMS} but for the resolved counterparts, $\Sigma_{*}$ and $\Sigma_{SFR}$. The spaxel-by-spaxel analysis is defined for a mass interval of approximately $6.87 \lesssim \log(\Sigma_{*}) \lesssim 8.33$, containing $80 \%$ of the data to exclude outliers. The OLS fitting results in the following fit: $b = 0.425 \pm 0.014$ and $a = -5.428 \pm 0.104$. The slope and zero-point via ODR fitting are: $b = 1.162 \pm 0.022$ and $a = -11.014 \pm 0.164$. The ODR results have a significantly steeper slope than via the OLS fitting method. We measure a $1\sigma$ scatter, derived by calculating the standard deviation of the residual between the observed and fitted $\log(\Sigma_{SFR})$, of $0.36$ dex.

\begin{table*}[]
    \centering
    \caption{Best-fit values of the rSFMS for our work and several other publications.}
    \begin{tabular}{c c c c c c}
    \hline
    \hline
    Reference & Data \tablefootmark{a}& Method \tablefootmark{b} & z & b & a \\
    \hline
    1 & 80\% & OLS &  0.3 &  $0.425 \pm 0.014$ & $-5.428 \pm 0.104$\\
    1 & 80\% & ODR & 0.3 & $1.162 \pm 0.022$ & $-11.014 \pm 0.164$ \\
    2 & 80\% & OLS & 0.3 &   $0.401 \pm 0.015$& $-5.615 \pm 0.116$ \\
    2 & 80\% & ODR & 0.3 & $2.562 \pm 0.056$ & $-22.248 \pm 0.431$ \\
    3 & 80\% & - &0 &  $0.72\pm 0.04$ & $-7.95\pm 0.29$ \\
    4 &- & OLS &  $<0.15$ & $0.715 \pm 0.001$ & $-8.056 \pm 0.008$ \\
    4 &- & ODR &  $<0.15$ & $1.005 \pm 0.004$ & $-10.338 \pm 0.014$ \\
    5 & $\log(\Sigma_{*})>7$&  OLS & 0.26 & $0.771 \pm 0.032$ & $-7.812 \pm 0.249$ \\
    6 & - & LTS & $0.25<z<0.42$ & $0.918 \pm 0.005$ & $-9.196 \pm 0.006$ \\
    7 & - & OLS& $0.1<z<0.42$ & $0.43\pm 0.05$ & $-5.66\pm 0.05$ \\
    8 & $\log(\Sigma_{*})<8.8$ & - & $0.7<z<1.5$ & $0.95$  & $-8.4$ \\
    \hline 
    \end{tabular}
    \tablefoot{Results by \cite{Mun_2024} also utilize the MAGPI survey but for a wider redshift range of $0.25<z<0.42$. 
    \tablefoottext{a}{Data range used for the fitting.}
    \tablefoottext{b}{Linear fitting method: ordinary least-square (OLS), orthogonal distance regression (ODR), or least trimmed squares (LTS).}}
    \tablebib{(1)~This work (SF-spaxels); (2)~This work (spectral decomposition); (3)~\citet{Cano_Diaz_2016_rSFMS}; (4)~\citet{Hsieh_2017_rSFMS}; (5)~\citet{Yao_2022}; (6)~\citet{Mun_2024}; (7)~\citet{Jafariyazani_2019_rSFMS}; (8)~\cite{Wuyts_2013_rSFMS}.}
    \label{tab:rSFMS_fits}
\end{table*}

Using EAGLE hydrodynamical cosmological simulations, \cite{Trayford_Schaye_2019_rSFMS_sim}, predict that the slope of the rSFMS increases with increasing redshift, which they attribute to inside-out galaxy formation. Table~\ref{tab:rSFMS_fits} lists our results and those from several other publications, showing that our resulting OLS slope diverges from the expected trend in rSFMS slope evolution with redshift. In Table~\ref{tab:rSFMS_fits} we also list each publication's utilized data range and fitting routine method.  However, the OLS slope aligns well with the results obtained from \cite{Jafariyazani_2019_rSFMS}, who utilized MUSE data at $0.1<z<0.42$ but for lower stellar masses. Nonetheless, our relatively shallow OLS rSFMS slope might result from MAGPI fields often probing dense environments, which are more prone to phenomena suppressing star formation such as environmental quenching (see e.g., \citealt{Mao_2022, Taylor_2023}). On the other hand, \cite{Popesso_2023_SFMS_cosmictime} recently conducted a study of the SFMS redshift evolution, where they compiled existing studies of the SFMS over a total redshift range of $0<z<6$ and stellar mass of $8.5 < \log(M_{*}/M_{\odot})<11.5$. They find that only the normalization and turnover mass evolve over time, but not the slope.

We also applied an ODR fitting routine, as shown in Table~\ref{tab:rSFMS_fits}, and the resulting slope aligns much better with the expected redshift evolution. \cite{Hsieh_2017_rSFMS} also analyzed the rSFMS using both an OLS and ODR fitting routine. They find that a steeper slope is measured via the ODR method. Comparing our ODR slope to that of \cite{Hsieh_2017_rSFMS}, we can detect a redshift evolution. 

The SFR values for our sample match those from \cite{Jafariyazani_2019_rSFMS} and are mainly located above the local Universe's SFRs from \cite{Cano_Diaz_2016_rSFMS}. This result agrees with the observations that galaxies at higher redshifts have higher SFRs than in the local Universe. Nonetheless, we observe lower SFRs than \cite{Yao_2022} who conducted a similar analysis using MUSE data at a redshift of $z\sim 0.26$. We conclude that this could be caused by the differences in the selection and depth of the data used and the S/N criteria applied to select spaxels.

\cite{Mun_2024} also investigate the rSFMS via MAGPI data using a least trimmed squares (LTS) routine and find a slope of 0.922, which aligns more with the findings of the redshift evolution than our OLS slope. Comparing their LTS results to our ODR slope, we are in relatively good agreement. Their results are also plotted as a dark-blue line in Fig.~\ref{fig:rSFMS}. However, their analysis was conducted within a redshift range of $0.25<z<0.42$ and a much wider total stellar mass range, including low-mass galaxies within $6.2 \lesssim \log(M_{*}/M_{\odot}) \lesssim 11.4$. As we need sufficient S/N in all four emission lines $[OIII] 5006 \AA$, $H\beta$, $H\alpha$, and $[NII] 6585 \AA$, and therefore applied an S/N cut when selecting our sample, our sample size is much smaller in comparison. The wider mass range could explain their higher completeness within the low-$\Sigma_{*}$ and low-$\Sigma_{SFR}$ section of the $\log \Sigma_{*}$ versus $\log \Sigma_{SFR}$ distribution compared to our analysis. 

We repeated the same analysis of the rSFMS by applying the spectral decomposition method introduced in section~\ref{sec:SDM} to all spaxels present in the BPT diagram and show the result as a pink-colored line in Fig.~\ref{fig:rSFMS}. For this, we computed the fractional contribution of SF, AGN, and LINER to the $H\alpha$ emission line and corrected our $H\alpha$ fluxes to only consist of its SF contribution. This method aims to have a larger sample available for investigating the rSFMS, as our normal attempts consist of filtering out any spaxels that fall above the empirical line from \cite{Kauffmann_2003}, which inherently limits our sample. The resulting OLS rSFMS has the following parameters: $a=-5.359  \pm 0.082$ and $b= 0.368 \pm 0.011$. Additionally, we repeated an ODR analysis and find the following results: $a=-22.248  \pm 0.431$ and $b= 2.562 \pm 0.056$. The OLS slope of this result is consistent with the OLS slope for SF spaxels, but the SFRs are located below the results for SF spaxels. We consider the difference in the zero-point of our two fits to stem from the fact that there are more low $\log(\Sigma_{SFR})$ spaxels via the spectral decomposition method compared to only selecting SF-spaxels.

\subsection{Resolved MZR}\label{sec:rMZR}

\begin{figure}
    \centering
    \includegraphics[width=\linewidth]{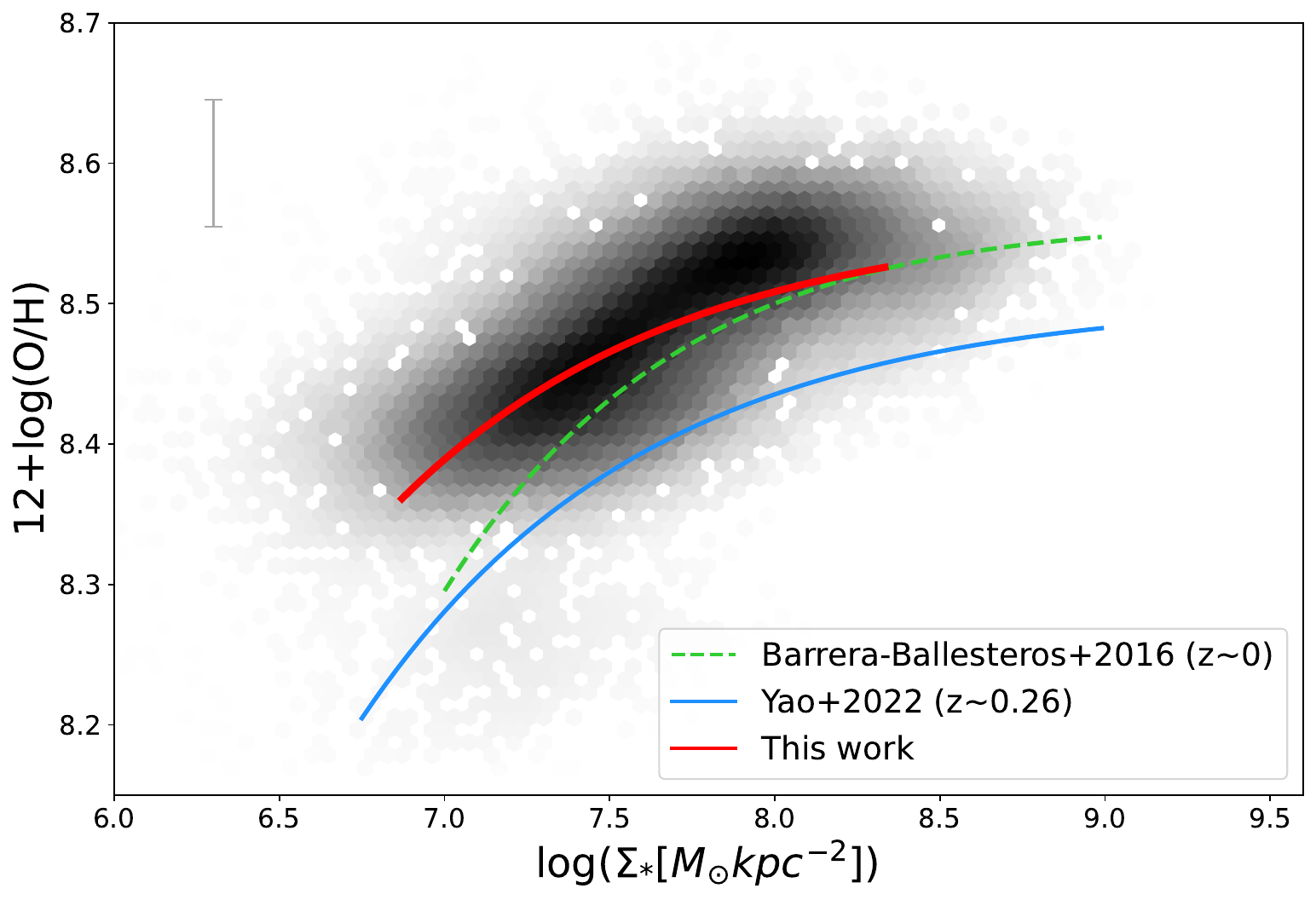}
    \caption{Distribution of the gas-phase metallicities of our sample of star-forming spaxels against their stellar mass surface densities. The hexagonal bins depict the sample distribution with darker colors corresponding to a denser region of data points. The corresponding fitting result for our data within a mass range of $6.87 \lesssim \log \Sigma_{*} \ (M_{\odot} kpc^{-2}) \lesssim 8.33$, which is defined as the range that covers $80 \%$ of the data, was computed via Eq.~\ref{eq:rMZR_fit} from \cite{Sanchez_2013_rMZR} and is shown as a red line. The gray errorbar in the top left corner shows the average uncertainties in gas-phase metallicity and $\Sigma_{*}$. FADO only gives errors based on the goodness of the fit, stellar mass errors are very small, and the average $\Sigma_{*}$ is not visible within this plot.}
    \label{fig:rMZR}
\end{figure}

Figure~\ref{fig:rMZR} shows the distribution of the gas-phase metallicity for star-forming spaxels as a function of their stellar mass surface density. The gray hexagonal bins represent our star-forming spaxels. As the surface mass density increases, the oxygen abundances also increase. As noted by previous studies (see e.g., \citealt{Tremonti_2004, Zahid_2013}), the relation flattens toward higher stellar masses. The average uncertainty of the gas-phase metallicity is $\sim 0.08$ dex. Within a mass range of $6.87 \lesssim \log \Sigma_{*} \ (M_{\odot} kpc^{-2}) \lesssim 8.33$, which is defined as the range that covers $80 \%$ of the data, we performed a nonlinear least square fit via the following equation from \cite{Sanchez_2013_rMZR}: 

\begin{equation}\label{eq:rMZR_fit}
    12+\log \left(O/H\right) = a + b \left(\log \Sigma_{*}-c\right) \exp{\left(-\left(\log \Sigma_{*}-c\right)\right)}.
\end{equation}

The results from our fitting routine for the coefficients are $a= 8.561 \pm 0.016$,  $b = 0.0001 \pm 0.0006$, and $c=12.7 \pm 5.542$. We obtain a $1\sigma$ scatter, derived by calculating the standard deviation of the residual between the observed and fitted gas-phase metallicity, of $0.08$ dex. 

We compare our results to the results obtained by \cite{Barrera_Ballesteros_2016_rMZR} for the local Universe utilizing 653 galaxies from the MaNGA survey covering a total stellar mass range of $8.5 \lesssim \log(M_{*}/M_{\odot}) \lesssim 11$ and those by \cite{Yao_2022} at a similar average redshift of $z\sim 0.26$ over a mass range of $9 \lesssim \log(M_{*}/M_{\odot}) \lesssim 10.6$ using the MUSE wide survey. \cite{Barrera_Ballesteros_2016_rMZR} also applied the same strong-line metallicity calibration by \cite{Marino_2013_metallicity} from Te-based HII regions, and differences in the IMF are accounted for. \cite{Yao_2022} utilized metallicity calibration from ONS-based HII regions by \cite{Marino_2013_metallicity} and we have adjusted for the difference between their ONS-based and our Te-based calibrations. Results by \cite{Yao_2022} are of a similar shape, but their metallicities are below ours, with an average downward shift of $\sim 0.09$ dex. This could be explained by their overall smaller sample, which also covers a smaller total mass range, whereas in our case, about a third of our sample exhibits $\log(M_{*}/M_{\odot}) > 10.6$, leading to our sample being more metal-rich. We, therefore, investigated whether our sample's $M_{*}$ range might affect our resulting oxygen abundances. We created a subset of galaxies within our sample that matches the mass range $9 \lesssim \log(M_{*}/M_{\odot}) \lesssim 10.6$ from \cite{Yao_2022}. Still, the resulting rMZR fit is on average at only $0.009$ dex lower than our entire sample. Thus, we cannot confirm that the differences in our gas-phase metallicities from those by \cite{Yao_2022} stem from any difference in the sample's total stellar mass range.

Gas-phase metallicities are observed to evolve with redshift: for a given stellar mass, metallicity declines with increasing redshift. This evolution is strongly defined within low-mass galaxies, while high-mass galaxies are believed to have already reached their local metallicity values at $z\sim 1$ due to downsizing \citep{Maiolino_2019_deremet}. \cite{Barrera_Ballesteros_2016_rMZR} results of the rMZR for the local Universe are situated below those of ours, with a $\sim 0.03$ dex downward shift, which again confirms the overall high metallicities within our sample. Therefore, we do not observe any redshift evolution of the rMZR. On the other hand, \cite{Barrera_Ballesteros_2016_rMZR} results are significantly above those obtained by \cite{Yao_2022} at $z\sim 0.26$, with the former aligning more reasonably with the expected metallicity redshift evolution. Furthermore, as mentioned in section~\ref{sec:MZR}, \cite{Maier_2015_MZR} found that there is only a low offset between the gas-phase metallicities at $z=0$ and at $0.5<z<0.75$. Assuming similarity in the resolved version, we expect only a small offset between our sample at $z=0.3$ and the local Universe. 

Lastly, we conducted a simple environmental analysis of the gas-phase metallicities within our sample. Gas-phase metallicities have consistently been observed to be higher in dense environments than in field galaxies (see e.g., \citealt{Cooper_2008}), which could explain why our values for MAGPI galaxies, which lie mostly within dense environments, are similar to those for the local Universe. To investigate this, we separated MAGPI's primary groups from field galaxies by considering the following selection: $|z_{secondary}-z_{primary}|<0.03$. Here, $z_{primary}$ are redshifts of MAGPI's group galaxies, while $z_{secondary}$ are redshifts of secondary galaxies within the same field as its corresponding primary galaxy. Galaxies with a redshift difference lower than $0.03$ are considered part of the primary group. We find that group galaxies have consistently higher gas-phase metallicities than field galaxies, averaging about $0.02$ dex. Also, the resulting rMZR fit for group galaxies is, on average, $0.01$  dex higher than for field galaxies. This is consistent with results from other works such as \cite{Geha_2024} which investigated the environmental processes of star-forming properties of 378 satellite galaxies in the local Universe from the Satellites Around Galactic Analogs (SAGA; \citealt{Geha_2017_SAGA}) survey and \cite{Maier_2022} who studied 18 clusters from the Local Cluster Substructure Survey (LoCuSS; \citealt{Smith_2010_Locuss}) at $z\sim 0.2$.

\subsection{Resolved FMR}

\begin{figure}
    \centering
    
    \includegraphics[width=\linewidth]{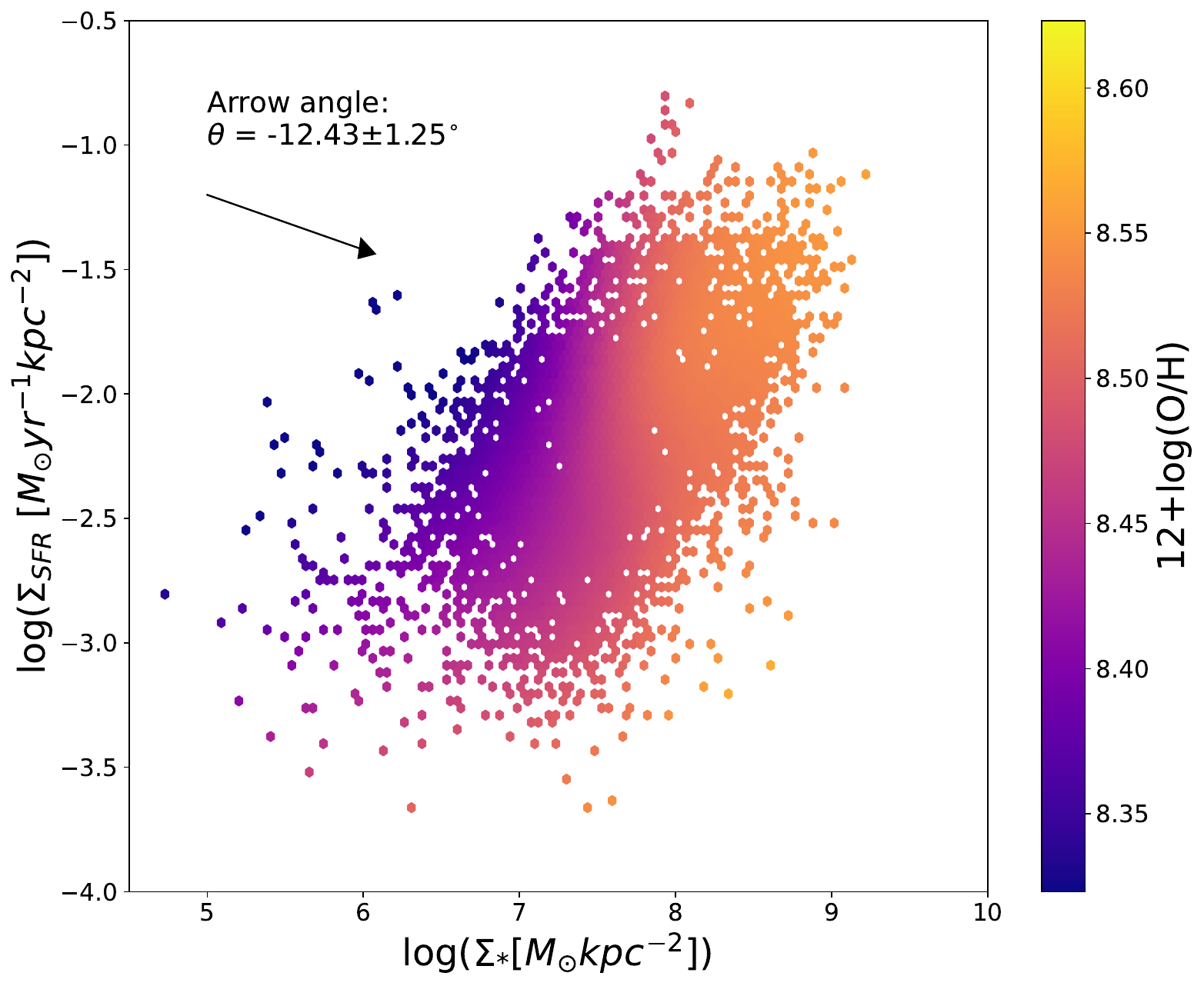}
    \caption{Resolved fundamental metallicity relation for the star-forming spaxels of our sample. Here, $\log(\Sigma_{*})$ is plotted against $\log(\Sigma_{SFR})$ and each spaxel is color-coded by its gas-phase metallicity $12+log(O/H)$. Additionally, an arrow angle and its error point to the direction of the steepest average gradient of increase of the local metallicity. Images are smoothed over via LOESS \citep{Cappellari2013b_Loess}, a locally weighted regression method to compute mean trends of the population from noisy data. }
    \label{fig:rFMR}
    
\end{figure}

To produce quantitative statistical results for our analysis of the spatially resolved FMR, we make use of a partial correlation coefficient (PCC) analysis similar to the one presented by \cite{Baker_2022_rSFMS_rSK_rMGMS, Baker2022_metallicity}. The analysis involves determining the partial correlation between two variables while keeping other variables constant. This approach helps us discern the genuine and inherent correlations between the two variables, distinguishing them from indirect correlations arising from other scaling relationships. In this sense, an equation can be defined where the partial correlation between two variables, $A$ and $B$, is tested while controlling for a third variable, $C$:
\begin{equation}
    \rho_{A B | C} = \frac{\rho_{AB}-\rho_{AC} \ \rho_{BC}}{\sqrt{1-\rho_{AC}^2} \ \sqrt{1-\rho_{BC}^2}}.
\end{equation}

Here, $\rho_{XY}$ refers to the Spearman rank correlation coefficient between two variables $X$ and $Y$. It is important to note that PCCs will only give meaningful results for monotonic relations. In the case of our analysis, it is helpful to picture the dependence of the three variables as a 3D space where the points on B (y-axis) are plotted versus C (x-axis) and color-coded by A (z-axis). In this sense, gas-phase metallicities are our variable A, $\Sigma_{SFR}$ is B, and $\Sigma_{*}$ is C. Another useful way to quantify PCC results is to produce an arrow pointing toward the steepest average gradient of increase of the variable A, in this case, the metallicity, therefore allowing us to determine the role of B and C, $\Sigma_{SFR}$ and $\Sigma_{*}$, in driving the gas-phase metallicity A. The angle $\Theta$ of this arrow is measured from the horizontal (three o'clock position) and is defined by the following equation from \cite{Bluck_2020_arrow}: 
\begin{equation}\label{eq:arrow_angle}
    \tan(\Theta) = \frac{\rho_{B A| C}}{\rho_{C A | B}}.
\end{equation}

Thus, this arrow angle is derived by taking the ratio of the PCC, looking at the influence of $\Sigma_{SFR}$ on the metallicity while controlling for $\Sigma_{*}$ and the PCC determining the influence of $\Sigma_{*}$ on the metallicity while controlling for $\Sigma_{SFR}$. To obtain arrow angle uncertainties, we applied the same method as \cite{Baker2022_metallicity} and utilized bootstrap random sampling to 100 random samples of the data and compute their standard deviation.

Figure~\ref{fig:rFMR} shows $\log(\Sigma_{SFR})$ versus $\log(\Sigma_{*})$, color-coding our data points with the local gas-phase metallicity $12+log(O/H)$ and including the arrow angle as defined in Eq.~\ref{eq:arrow_angle} in the top left corner of each Figure. This figure is smoothed using LOESS, a Python package that applies a locally weighted regression method to obtain mean trends of noisy data \citep{Cappellari2013b_Loess}. The stronger the arrow points vertically, the stronger the influence of $\Sigma_{SFR}$ on driving the local metallicities. Similarly, the more horizontal the arrow appears, the stronger the influence of $\Sigma_{*}$ in determining the local gas-phase metallicity.

The color shading alone demonstrates how $\Sigma_{SFR}$ and $\Sigma_{*}$ drive the local metallicity. If only $\Sigma_{*}$ influenced the metallicity, there would be vertical gradients of color as the metallicity would not change with $\Sigma_{SFR}$ at all. On the contrary, if only $\Sigma_{SFR}$ determined the metallicity, we would expect results of horizontal gradients of color. In our case, we see a color shading that indicates that the metallicity is proportionally correlated with $\Sigma_{*}$ and inversely correlated with $\Sigma_{SFR}$. The direction of the color shading also indicates that $\Sigma_{*}$ is the dominant factor in determining the metallicity. 

We look at the arrow angle depicted in Fig.~\ref{fig:rFMR} to better understand these correlations. For our sample, we obtain a result of $\Theta=-12.43 \pm 1.25^{\circ}$. This corresponds to a $13.81\%$ contribution from $\Sigma_{SFR}$ and $86.19\%$ contribution from $\Sigma_{*}$ to the spatially resolved gas-phase metallicity. The direction in which the arrow angle points indicates that $\Sigma_{*}$ has a stronger correlation with the gas-phase metallicity than $\Sigma_{SFR}$ does, but that both combined determine the metallicity. In conclusion, there is a small but significant contribution from $\Sigma_{SFR}$ in driving the local metallicity, although $\Sigma_{*}$ remains the dominant influence. Therefore, to increase the metallicity, an increase in $\Sigma_{*}$ is needed, which is what would be expected from the rMZR. Still, a secondary inverse dependence exists on $\Sigma_{SFR}$ for the metallicity. An increase in $\Sigma_{SFR}$ weakly correlates with a decrease in the local metallicity. Our ratio is much smaller than the one reported by \cite{Baker2022_metallicity} of $\Theta = -40^{\circ}$ for their entire sample of local galaxies covering stellar masses of $9.0<\log(M_{*}/M_{\odot})<11.4$.

We repeated our analysis of the arrow angle by grouping our data into different bins of total stellar mass. The results are summarized in Table~\ref{tab:arrow_angles}. In conclusion, galaxies at the lower-mass end of our sample show a significantly stronger contribution of $\Sigma_{SFR}$ on the metallicity than higher-mass galaxies. \cite{Baker2022_metallicity} also repeated their analysis by grouping their data into three separate mass bins, with their highest mass bin of $10.6<\log(M_{*}/M_{\odot}) <11.4$ resulting in an arrow angle of $-16.6^{\circ}$ and, therefore, they report a ratio of $18.44\%$ $\Sigma_{SFR}$ and $81.56\%$ $\Sigma_{*}$ contribution. Hence, their highest mass bin arrow angle aligns the most with our results. Generally, their results indicate a stronger influence of $\Sigma_{SFR}$ on the metallicity in the local Universe than our overall results indicate at $z\sim 0.3$.

\begin{table*}[]
    \centering
    \caption{Results for the arrow angle and the corresponding percentage contributions of Eq.~\ref{eq:arrow_angle} for different mass bins.}
    \begin{tabular}{c c c c c}
    \hline
    \hline
    Mass bin & \# Galaxies & Arrow angle & $\Sigma_{SFR}$ [\%] & $12+\log(O/H)$ [\%] \\
    \hline
    all & 65 &  $-12.4 \pm 1.25^{\circ}$ & 14 & 86\\
    $ \log(M_{*}) < 9$ & 12 & $-65.9 \pm 6.8^{\circ}$ & 73 & 27\\
    $9.0 < \log(M_{*}) < 9.8$ & 16 &  $-38.6 \pm 2.2^{\circ}$ & 43 & 57\\
    $9.8 < \log(M_{*}) < 10.6$  & 17 & $-12 \pm 2.3^{\circ}$ & 13 & 87\\
    $10.6< \log(M_{*}) < 11.4$  & 20 & $1.5 \pm 2.3^{\circ}$ & 2 & 98\\
    \hline
    \end{tabular}
    \label{tab:arrow_angles}
\end{table*}

\begin{figure}
    \centering
    \includegraphics[width=\linewidth]{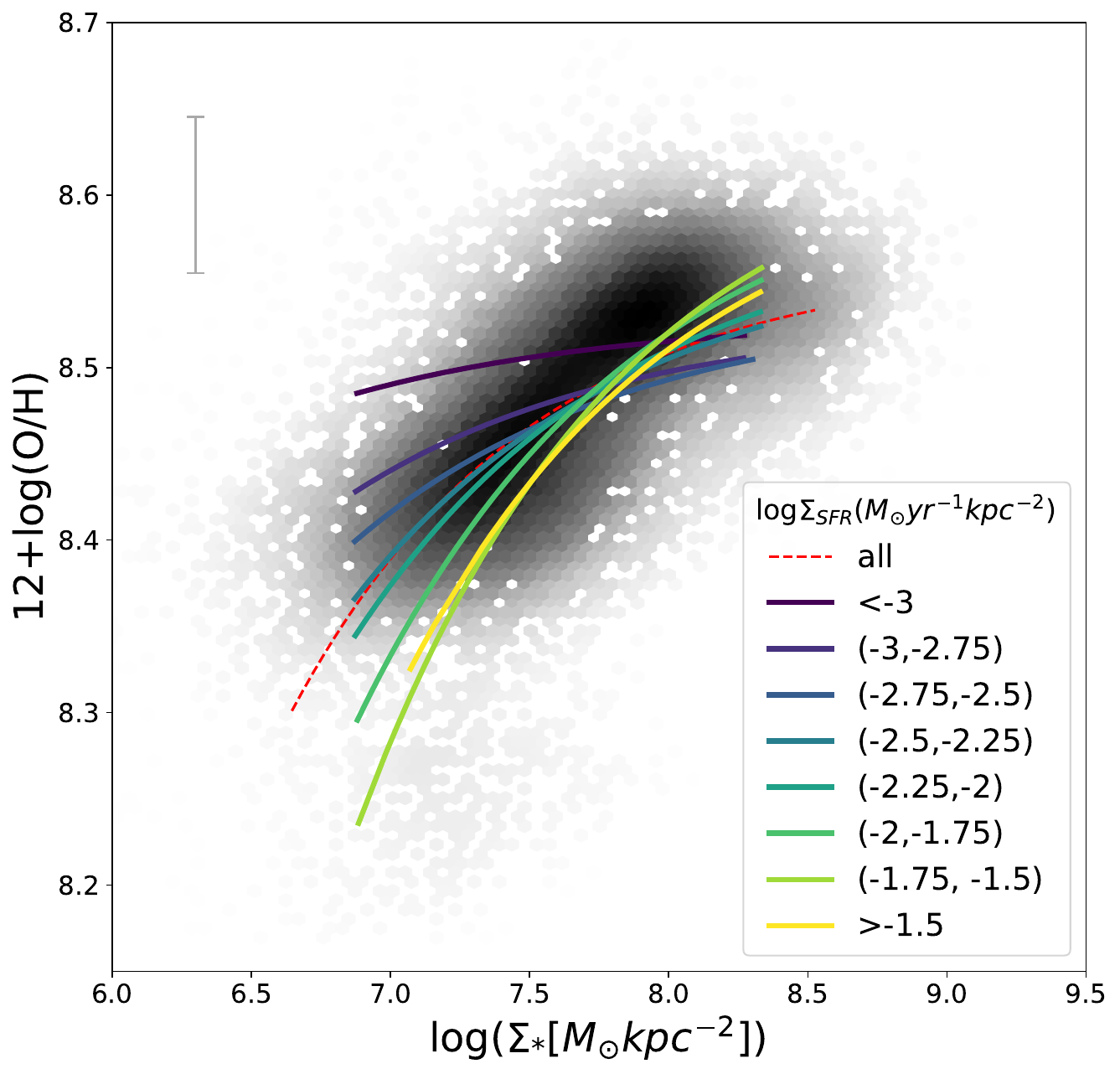}
    \caption{Resolved mass-metallicity relation for different $\Sigma_{SFR}$ bins, as indicated in the legend. Each line represents a rMZR fit for the different corresponding $\Sigma_{SFR}$ bins computed as described in section~\ref{sec:rMZR}. The dashed red line represents the results for all SF spaxels following Eq.~\ref{eq:rMZR_fit}. The gray errorbar shows the average uncertainties in gas-phase metallicity and $\Sigma_{*}$. FADO only gives errors based on the goodness of the fit, stellar mass errors are very small, and the average $\Sigma_{*}$ is not visible within this plot.}
    \label{fig:rFMR_SFRbins}
\end{figure}

Figure~\ref{fig:rFMR_SFRbins} shows the rMZR binned by $\Sigma_{SFR}$, similar to the FMR analysis done by \cite{Mannucci_2010}, and is a different way to visualize the rFMR. Here, the different colored tracks correspond to the rMZR fits done via Eq.~\ref{eq:rMZR_fit} for different $\log(\Sigma_{SFR})$ bins, as described in the legend, while the dashed red line is the rMZR fit derived in section~\ref{sec:metals} covering the entire sample. There were substantial outliers toward lower stellar mass surface densities, so the different bins are plotted within a $\log(\Sigma_{*})$ range that covers $80\%$ of the data. Figure~\ref{fig:rFMR_SFRbins} further supports the existence of an rFMR for lower $\Sigma_{*}$. At roughly $\log(\Sigma_{*})\sim 7.8$, the inverse correlation between gas-phase metallicity and SFR flattens out and even seemingly inverts to a certain degree. At any point below $\log(\Sigma_{*})\sim 8$, we find that for a fixed $\Sigma_{*}$, higher $\Sigma_{SFR}$ correlate with lower $12+log(O/H)$.

\begin{figure}
    \centering
     \includegraphics[width=\linewidth]{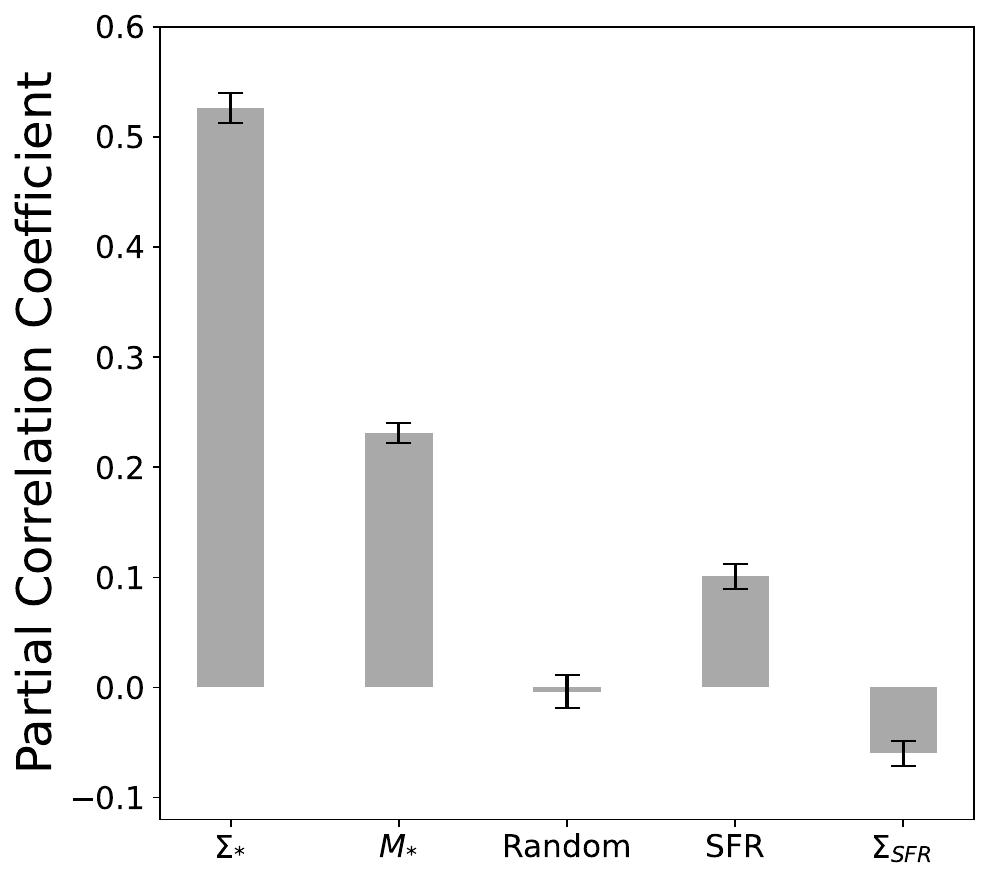}
    \caption{Partial correlation coefficients of the local metallicity and the resolved stellar mass surface density $\Sigma_{*}$, total stellar mass $M_{*}$, a uniform random variable, resolved star formation rate $\Sigma_{SFR}$, and total star formation rate SFR. Errors shown as error bars are computed via bootstrap random sampling.}
    \label{fig:PCCs}
\end{figure}

\subsection{Local metallicity dependence on resolved and global properties}

To get a deeper look at the strengths of the relations defining the local metallicity, we computed the partial correlation coefficients between the resolved metallicity and $\Sigma_{*}$, $M_{*}$, $\Sigma_{SFR}$, and SFR while controlling for either $\Sigma_{*}$ (when analyzing the correlation with $M_{*}$), $M_{*}$ (when analyzing the correlation with $\Sigma_{*}$), or both (when analyzing the correlation with $\Sigma_{SFR}$ and $SFR$). 

Results are shown as a bar chart in Fig.~\ref{fig:PCCs} with error bars indicating the uncertainties obtained via bootstrap random sampling. A uniform random variable was also included as a control mechanism. This analysis provides us with the strength and sign of the correlations, although it should be mentioned again that PCCs only work accurately for monotonic relations. The local stellar mass density $\Sigma_{*}$ is the main driver of the local metallicity, followed by the total stellar mass $M_{*}$. The total SFR and $\Sigma_{SFR}$ show only a weak correlation, with the total SFR even positively correlating with the local metallicity. The strength at which $\Sigma_{SFR}$ negatively correlates with the gas-phase metallicity is also much weaker than the results from \cite{Baker2022_metallicity} for the local Universe.

We note that \cite{Baker2022_metallicity} also employed a random forest algorithm to determine the relative importance of each parameter on the metallicity, which has better accuracy in determining their true correlations than the Partial Correlation Coefficient analysis does. What is important to conclude from these results is that $\Sigma_{*}$ is evidently the dominating influence on the local metallicity, followed by total stellar mass $M_{*}$. Suppose the partial correlation coefficient analysis should only be taken at face value regarding the sign of the correlations and not at their correlation strength. In that case, these results indicate an anti-correlation between $\Sigma_{SFR}$ and $12+\log(O/H)$.

To give another way to quantify our results of the rFMR, we investigated the relation between the gas-phase metallicity and the parameter $\mu_{\alpha}$, which was first defined by \cite{Mannucci_2010}: 
\begin{equation}\label{eq:mu_alpha}
     \mu_{\alpha} = \log{\Sigma_{*}} - \alpha \cdot \log{\Sigma_{SFR}}. 
\end{equation}

This function is defined to parametrize the projection of the FMR that minimizes the scatter in metallicity. Therefore, $\alpha = 0$ would correspond to $\mu_0 = \log\left(M_{*}\right)$, while $\alpha = 1$ would result in $\mu_1 = - \log\left(sSFR\right)$. Their argument for this type of parametrization is that the value of $\alpha$ that minimizes the scatter of the median metallicities around the relation corresponds to a $\mu_{\alpha}$ that exhibits the tightest possible, "most fundamental" correlation with the gas-phase metallicity.

\begin{figure}
    \centering
    \includegraphics[width=\linewidth]{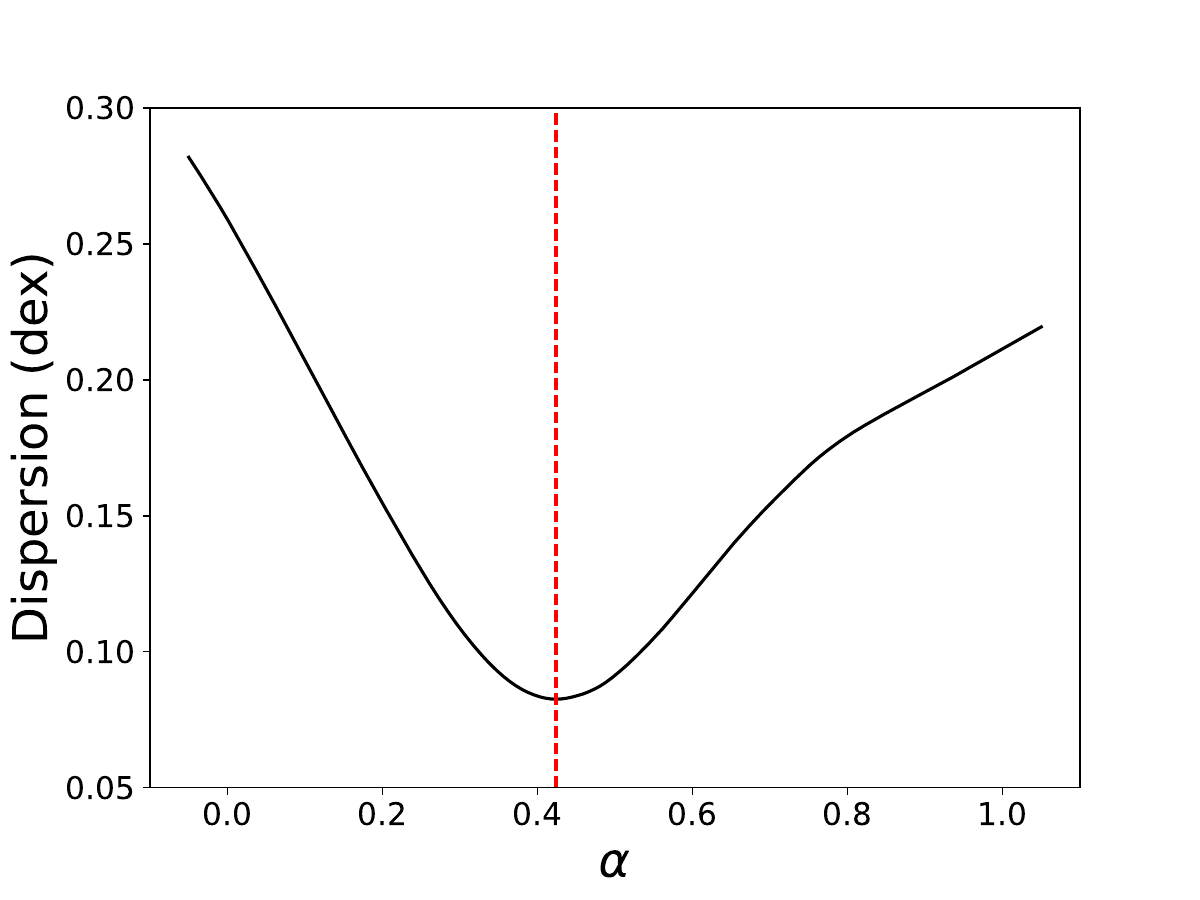}
    \caption{Residual dispersion of the gas-phase metallicities as a function of $\alpha$ following Eq.~\ref{eq:mu_alpha}. The value of $\alpha = 0.42$, corresponding to the minimum dispersion, is highlighted as a dashed red line.}
    \label{fig:alpha}
\end{figure}

Here, the function $\mu_{\alpha}$ is minimized to find the value of $\alpha$ that minimizes the dispersion between $12+log(O/H)$ and $\mu_{\alpha}$. For our sample, we find $\alpha=0.42$ for SF spaxels with an average dispersion of around $\sim0.09$ dex, around $\sim 0.17$ dex lower than the average dispersion for $\alpha=0$, confirming the existence of an rFMR. We plot the dispersion as a function of $\alpha$ in Fig.~\ref{fig:alpha} and indicate our minimum dispersion as a dashed red line. Our reported $\alpha$ parameters also align well with the initial finding of $\alpha = 0.32$ by \cite{Mannucci_2010}, which results in an average dispersion that is only $0.02$ dex higher than that of our optimal $\alpha$. Other investigations into the spatially resolved $12 + \log O/H$ versus $\mu_{\alpha}$ relation have resulted in various values: \cite{Baker2022_metallicity} ($\alpha = 0.54$, $z\sim 0$), \cite{Li_2024} ($\alpha = 0.33$, $0.01 \lesssim z \lesssim 0.15$), \cite{Andrew_Martini_2013_FMR_alpha} ($\alpha = 0.66$, $0.027 <z<0.25$), and \cite{Yao_2022} ($\alpha = 0.51$, $z \sim 0.26$).

\subsection{Voronoi binning}

We also repeated our entire analysis via Voronoi binning (Vorbin; \citealt{Capellari_2003_Vorbin}), binning our spaxel data to ensure a high enough S/N of the emission lines for our spatially resolved analysis of gas-phase metallicities. This method relies on Voronoi tessellations to bin spectral data to a minimum S/N requirement. 

Vorbin needs a target S/N value for each spaxel. We computed the S/N within an observed wavelength window of $6050\AA < \lambda_{obs} < 7750\AA$, corresponding roughly to the peak MUSE sensitivity range, and set the minimum target S/N to 8. Only spaxels with a minimum S/N of 2 were included in the binning process. All spectra in one bin were stacked and averaged. We applied the same steps regarding extinction correction and emission line flux measurements as described in section~\ref{sec:methods}. To conclude, we do not find significant differences between our results via a spaxel-by-spaxel analysis and Voronoi binning.

\section{Discussion}\label{sec:discussion}

Our main goal is to determine the interdependence between the spatially resolved and global parameters defining the rFMR. We find that for our sample, the local gas-phase metallicity primarily correlates with $\Sigma_{*}$ and has an inverse secondary correlation with $\Sigma_{SFR}$, as can be seen per Fig.~\ref{fig:rFMR} and \ref{fig:rFMR_SFRbins}, confirming an rFMR at $z\sim 0.3$. This aligns well with the theory that the FMR is caused by the accretion of pristine metal-poor gas, which both dilutes the overall gas content in a galaxy, decreasing its metallicity and increases the overall SFR as it acts as a star formation boost (see e.g., \citealt{Tremonti_2004, Mannucci_2010, Lara-Lopez_2010_FMR, Lilly_2013}). However, the fact that the global SFR has a stronger and, more importantly, positive correlation with the local gas-phase metallicity majorly diverges from this theory. These results also greatly contrast what \cite{Baker2022_metallicity} find using the same partial correlation coefficient analysis for the local Universe, where their inverse correlation with both SFR and $\Sigma_{SFR}$ underlines the necessity of both the accretion-dilution scenario that is typically associated with the FMR and the need for metal-rich galactic-scale outflows caused by winds. 

When analyzing the rFMR, we only find a weak correlation between the local metallicity and $\Sigma_{SFR}$ as indicated by the arrow angle. Our angle of $-12.43^{\circ}$ corresponds to a roughly $14 \%$ contribution from $\Sigma_{SFR}$ and $86\%$ contribution from $\Sigma_{*}$ to the spatially resolved metallicity. However, when investigating these interdependences in more detail, we notice a significant turnover at $\log(\Sigma_{*}) \gtrsim 7.8$, where the rFMR starts to weaken and even inverts. This is best depicted in Fig.~\ref{fig:rFMR_SFRbins} where we plotted and fitted the rMZR for different $\Sigma_{SFR}$ bins. Similarly, \cite{Baker2022_metallicity} investigated the rFMR by plotting the metallicity versus $\Sigma_{*}$ binned by tracks of different $\Sigma_{SFR}$ and observed a similar inversion at high $\Sigma_{*}$. Furthermore, we also tested how well the rFMR is defined for a $\log(\Sigma_{*})<7.8$. We find that it produces an arrow angle of $\sim -27.3 \pm 1.5^{\circ}$, which indicates a stronger inverse influence of $\Sigma_{SFR}$ on the local metallicity than when considering our full $\Sigma_{*}$ range.

In section~\ref{sec:rMZR} we investigated whether our differences in rMZR between our sample and those by \cite{Yao_2022}, who also utilized MUSE data at a similar redshift of $z\sim 0.26$ but for a smaller stellar mass range of $9 \lesssim \log(M_{*}/M_{\odot}) \lesssim 10.6$, could be explained by the influence of $M_{*}$ on the rMZR. To conclude, we created a subsample of our data that matches the mass range of \cite{Yao_2022} and re-fitted the rMZR. The resulting fit is located only minimally below that for our entire sample, and we can not explain this difference via a dependence on $M_{*}$. However, as mentioned in section~\ref{fig:rMZR}, a difference in the exposure times of the observations between the MUSE-wide survey utilized by \cite{Yao_2022}, which has a total integration time of 1h and our MAGPI data with an exposure time of 4.4h per field, could possibly explain their overall higher SFRs and lower gas-phase metallicities compared to ours. 

Nevertheless, we note that a cut in $\Sigma_{*}$, as discussed above, is similar to a cut in $M_{*}$, as is indicated by the arrow angle results shown in Table~\ref{tab:arrow_angles} for different $M_{*}$ bins. Therefore, we find evidence of a significant influence of $M_{*}$ on the gas-phase metallicity via the partial correlation coefficients and arrow angle analysis. In fact, the correlation between $M_{*}$ and the local metallicity is stronger than between the metallicity and $\Sigma_{SFR}$. Only selecting low $\Sigma_{*}$ spaxels or low $M_{*}$ galaxies results in a stronger contribution from SFR to the spatially resolved metallicity. We also observe that the highest $\Sigma_{*}$ spaxels stem from our highest $M_{*}$ galaxies. Hence, we find good evidence for an rFMR at $z\sim 0.31$ for selecting spaxels with $\log(\Sigma_{*}) \lesssim 7.8$ or galaxies with $\log(M_{*}/M_{\odot}) \lesssim 10.6$. Beyond those, the relation starts to flatten out and even invert. Hence, the overall small contribution from $\Sigma_{SFR}$ toward the local metallicity for our entire sample could also be explained by our sample's bias toward higher $M_{*}$. 

\cite{Bulichi_2023}, who analyzed the rFMR for nine local dwarf galaxies via MUSE data, also concluded that the anti-correlation between gas-phase metallicity and $\Sigma_{SFR}$ is strongest for low-mass galaxies. This is also confirmed by the results from \cite{Curti_2019_MZR}. The fact that the FMR is most pronounced toward low-mass galaxies aligns well with our results and has even been replicated via EAGLE simulations \citep{Scholz-Diaz_2021}. This also follows expectations that SN feedback, which is responsible for metal supply, mixing, and ejection of metal-rich material and dilution effects caused by a strong inflow of metal-poor gas, are most prominent in low-mass galaxies. \cite{Mannucci_2010} argued that the outflow of gas is responsible for the metallicity's dependence on stellar mass, as winds are more efficient in carrying out metals inside lower-mass galaxies and that toward higher-mass galaxies, the efficiency of outflows decreases due to the galaxies larger potential wells. Therefore, both effects can decrease gas-phase metallicities in low-mass galaxies and fuel their star formation. Indeed, observations have shown that SN feedback-driven outflows exhibit higher gas-phase metallicities than the ISM of its origin galaxy \citep{Chisholm_2018}. On the other hand, AGN feedback starts to dominate galactic outflows in high-mass galaxies, which are generally better at retaining metals due to their deeper gravitational potential, thus decreasing the impact of $\Sigma_{SFR}$ on the local metallicity at higher $M_{*}$.

\section{Summary and conclusion}\label{sec:conclusion}

We utilized data from the MUSE Large Program MAGPI. We selected 65 galaxies with strong optical emission lines over a redshift range of $0.28<z<0.35$ and within a total stellar mass range of $8.2 < \log(M_{*}/M_{\odot}) < 11.4 $. We measured emission line fluxes and stellar masses via the population spectral synthesis code FADO to reliably determine the galaxies $\Sigma_{SFR}$, $\Sigma_{*}$, and spatially resolved gas-phase metallicities. 

Our main results are as follows: 
\begin{itemize}
    \item Via diagnostic diagrams, we confirm that the main ionizing mechanism in our sample is SF and that the empirical and theoretical line calibrations used to distinguish between the different ionizing mechanisms align well with the fractional contribution of SF, AGN, and LINER emission, computed via the spectral decomposition method from \cite{Davies17}, to the $H\alpha$ line. 
    \item We establish a global SFMS and MZR for our sample. The SFMS aligns well with the results found in other works. Our sample exhibits overall high gas-phase metallicities, as they are within the relation found for the local Universe.
    \item We confirm the existence of an rSFMS at $ z \sim 0.3$ with a slope of $ \sim 0.425$. This resulting slope is shallower than expected if a redshift evolution of the slope is assumed. However, via ODR fitting, we find a slope of $\sim 1.162$, which aligns better with the expected redshift evolution. Our spatially resolved SFRs are higher than those observed for the local Universe, which aligns with the theory that galaxies exhibit increasing SFRs toward higher redshifts. We find similar rSFMR slopes for our fits obtained by only selecting SF-spaxel, and when using the $H\alpha$ emission line, corrected to only include its SF contribution via the spectral decomposition method.
    \item The rMZR exists at $z\sim 0.3$, but we find overall high gas-phase metallicities which are on average $0.03$ dex higher than the local Universe's values, which is still within the diagnostic's uncertainty. This is also confirmed by our global MZR shown in Fig.~\ref{fig:global_relations}, where our galaxies are observed to be within the local relation. 
    \item We tentatively confirm the existence of an rFMR: the spatially resolved gas-phase metallicity primarily depends on the local stellar mass surface density $\Sigma_{*}$ with a secondary, inverse, dependence on the local star formation rate surface density $\Sigma_{SFR}$. 
    \item We investigated the correlation between global and resolved parameters through a partial correlation coefficient analysis. The two dominant factors in determining the local metallicity are $\Sigma_{*}$ and $M_{*}$, albeit, $\Sigma_{*}$ has a significantly higher contribution than $M_{*}$. Additionally, $\Sigma_{SFR}$ weakly inversely correlates with the local metallicity. We observe that the inverse correlation between the spatially resolved metallicity and $\Sigma_{SFR}$ significantly strengthens toward lower $M_{*}$ and $\Sigma_{*}$. 
\end{itemize}

\begin{acknowledgements}
We wish to thank the ESO staff, and in particular the staff at Paranal Observatory, for carrying out the MAGPI observations. MAGPI targets were selected from GAMA. GAMA is a joint European-Australasian project based around a spectroscopic campaign using the Anglo-Australian Telescope. GAMA was funded by the STFC (UK), the ARC (Australia), the AAO, and the participating institutions. GAMA photometry is based on observations made with ESO Telescopes at the La Silla Paranal Observatory under programme ID 179.A-2004, ID 177.A-3016. The MAGPI team acknowledge support by the Australian Research Council Centre of Excellence for All Sky Astrophysics in 3 Dimensions (ASTRO 3D), through project number CE170100013. CF is the recipient of an Australian Research Council Future Fellowship (project number FT210100168) funded by the Australian Government. CL, JTM and CF are the recipients of ARC Discovery Project DP210101945. SMS acknowledges funding from the Australian Research Council (DE220100003). LMV acknowledges support by the German Academic Scholarship Foundation (Studienstiftung des deutschen Volkes) and the Marianne-Plehn-Program of the Elite Network of Bavaria. KG is supported by the Australian Research Council through the Discovery Early Career Researcher Award (DECRA) Fellowship (project number DE220100766) funded by the Australian Government. PP acknowledges support by the Funda\c{c}\~{a}o para a Ci\^{e}ncia e a Tecnologia (FCT) grants UID/FIS/04434/2019, UIDB/04434/2020, UIDP/04434/2020 and Principal Investigator contract CIAAUP-092023-CTTI

\\

\\

This work uses the following software packages: Astropy \citep{Astropy}, FADO \citep{Gomoes_2017_FADO}, GIST \citep{Bittner_2019_GIST}, lmfit \citep{Newville_2016_LMFIT}, LOESS \citep{Cappellari2013b_Loess}, Matplotlib \citep{Hunter_2007_plt}, MPDAF \citep{Bacon_2016_MPDAF}, Numpy \citep{Harris_2020_numpy}, Pingouin \citep{Vallat2018_Punguoin}, Profound \citep{Profound_Robotham_2018}, Prospect \citep{Robotham_2020_prospec}, Scipy \citep{Virtanen_2020_scipy}, Seaborn \citep{Waskom_2021_seaborn}, Uncertainties\footnote{\url{https://pythonhosted.org/uncertainties/}}, and Vorbin \citep{Capellari_2003_Vorbin}.
    
\end{acknowledgements}

\bibliographystyle{aa}
\bibliography{bib.bib}

\end{document}